\documentclass[a4paper,fleqn,usenatbib]{mnras}
\usepackage[T1]{fontenc}
\usepackage{ae,aecompl}
\usepackage{natbib}
\usepackage{lscape}
\usepackage{graphicx}	% Including figure files
\usepackage{amsmath}	% Advanced maths commands
\usepackage{amssymb}	% Extra maths symbols
  \usepackage{multicol}        % Multi-column entries in tables
   \usepackage{bm}		% Bold maths symbols, including upright Greek
   \usepackage{pdflscape}	% Landscape pages
    \usepackage{subfig}
 \usepackage{enumerate}
 \usepackage{rotating}
\usepackage{float}
\usepackage{color}

\newcommand{\ha}{\mathrm{H}\alpha}
\newcommand{\hb}{\mathrm{H}\beta}

\newcommand{\hii}{H\,\textsc{II}}

\newcommand{\mincir}{\raise-3.truept\hbox{\rlap{\hbox{$\sim$}}\raise4.truept\hbox{$<$}\ }}

\newcommand{\rh}{R_{\rm h}}

\newcommand{\chifit}{\chi_t}
\newcommand{\gyr}{{\rm Gyr}}
\newcommand{\Msol}{M$_\odot$}
\newcommand{\msun}{M_\odot}

\newcommand{\myr}{{\rm Myr}}
\newcommand{\pc}{{\rm pc}}
\newcommand{\kpc}{{\rm kpc}}
\newcommand{\rhn}{R_{\rm h0}}

\newcommand{\mh}{M_{\rm h}}
\newcommand{\rg}{R_{\rm G}}
\newcommand{\rhoh}{\rho_{\rm h}}
\newcommand{\trh}{T_{\rm Rh}}
\newcommand{\trhn}{T_{\rm Rh0}}

\newcommand{\ts}{t_*}

\newcommand{\kms}{ km\ s$^{-1}$}                            % km s-1  
\newcommand{\ergs}{erg s$^{-1}$}                            %erg s-1  
\newcommand{\Msolar}{M$_{\odot}$}                           % Msolar   
\title[The evolution of Super Star Clusters]{
From Giant \hii\ regions and \hii\ galaxies to globular clusters and compact dwarf ellipticals.\\
} 

\author[E. Terlevich et al.]
{Elena Terlevich$^{1}$\thanks{E-mail: eterlevi@inaoep.mx},  David Fern\'andez-Arenas$^{1}$,
   Roberto Terlevich$^{1,2}$,  Mark Gieles$^{3}$, \newauthor Ricardo Ch{\'a}vez$^{1,4,5}$,
    and Ana Luisa Gonz\'alez-Mor\'an$^{1}$
  \\ \\
$^{1}$ Instituto Nacional de Astrof{\'i}sica {\'O}ptica y Electr{\'o}nica, AP 51 y 216, 72000, Puebla, M{\'e}xico.\\
$^{2}$ Institute of Astronomy, University of Cambridge, Madingley Road, Cambridge CB3 0HA, U.K.\\
$^{3}$ Department of Physics, University of Surrey, Guildford GU2 7XH, UK\\
$^{4}$ Cavendish Laboratory, University of Cambridge, 19 J. J. Thomson Ave, Cambridge CB3 0HE, UK.\\
$^{5}$  Kavli Institute for Cosmology, University of Cambridge, Madingley Rd., Cambridge, CB3 0HA, U.K.   \\
}

\begin{document}

%\date{v0.7b ET ---- Compiled at \thistime\ hrs  on \today\  }

%\pagerange{\pageref{firstpage}--\pageref{lastpage}} \pubyear{2014}

\maketitle

\label{firstpage}

\begin{abstract}

Massive starforming regions  like Giant \hii\ Regions (GHIIR) and \hii\ Galaxies (HIIG) are  emission line systems ionized by  compact young massive star clusters  (YMC) with masses ranging  from $10^4 \msun$  to $10^8 \msun$. 

We model the photometric and dynamical evolution over a Hubble time of  the massive gravitationally bound  systems that populate the tight relation between absolute blue magnitude and velocity dispersion ($M_{B}-\sigma$) of GHIIR and   HIIG and compare the resulting relation  with that one of  old stellar systems: globular clusters, elliptical galaxies, bulges of spirals.  
After 12~\gyr\ of evolution their position  on the $\sigma$ vs. M$_B$ plane  coincides -- depending on the initial mass  -- either with the globular clusters for systems with initial mass $M < 10^6 \msun$ or with a continuation of the ellipticals, bulges of spirals and ultracompact dwarfs for  YMC with $M >10^6 \msun$. 

The  slope change in the $M_{B}-\sigma$ and $M_B$-size relations at cluster masses around $10^6 \msun$ is due to the  larger impact of the   dynamical evolution on the lower mass clusters.

We interpret our result as an indication  that the YMC that ionize GHIIR and HIIG can evolve to form globular clusters and ultra compact dwarf ellipticals in about 12\,\gyr\ so that 
present day globular clusters and ultra compact dwarf ellipticals may have   formed in conditions similar to those observed  in today GHIIR and HIIG.

\end{abstract}


\section{Introduction}

The possible connection between young massive clusters (YMCs) and globular clusters (GCs) has been discussed in the literature mostly in relation with YMC found 
in the central galaxy of the Perseus cluster [NGC~1275; \citep{SF90, Holtzman92}]
in the central regions of interacting galaxies  \citep{Portegies2010}  and in local group galaxies like the Large and Small Magellanic Clouds.

The comparable masses and sizes of YMC and  GC lead to the belief that there might be  an evolutionary connection between these massive clusters posing the logical question of whether YMC could be  young GC.

It is important to note that while YMC found in massive interacting galaxies and GC have comparable masses, the former (unlike GC) are metal rich and are formed in high density environments. 

On the other hand, the YMC that ionize the GHIIR in dwarf irregular galaxies and  in the outer regions of late spirals,  have a range of masses and $\alpha$-element abundances (oxygen, neon, sulphur, argon) substantially subsolar,  similar to those found in GCs. In particular there have been suggestions that  30-Dor, the prototypical GHIIR in the LMC, could be a GC progenitor given its mass, size and metal content \citep[e.g.][]{Meylan1993}. 

Fe abundance is rarely measured in HII regions, as the ionization correction fractions for Fe can cause uncertainties of factors 2.5 - 3 depending on the degree of ionization of the region (M\'onica Rodr\'\i guez, private communication)  and the abundance may also be masked by depletion in dust grains \citep[e.g.][]{Esteban98} . In globular clusters, instead, the stellar abundances of $\alpha$-elements are normally given with respect to Fe ($\alpha$/Fe), and their value relative to H is then deduced from the Fe/H value.

The most extreme cases of low metallicity YMC are found in HIIG. These are YMC in dwarf galaxies that completely dominate the luminosity output having metallicities [O/H] down to 1/50th of solar [see e.g. the review by \cite{Kunth2000}].  
We have found \citep{Chavez2014, Terlevich2015}  that the properties of low-  and high-z HIIG are similar in every parameter that we have measured (mass, velocity dispersion, luminosity) thus strongly suggesting that the study of low {\it z}  YMC  can provide important clues to the formation and evolution of GCs. A particular case is that of ID11 an actively star-forming, extremely compact galaxy and  Ly$\alpha$  emitter at $z = 3.1$ that is gravitationally magnified by a  factor of $\sim$17 by the cluster of galaxies Hubble Frontier Fields AS1063. The size, luminosity, velocity dispersion and dynamical mass of ID11 resemble those of low luminosity HIIG or GHIIR such as 30 Doradus \citep{Terlevich2016}.

HIIG are  narrow emission line compact starforming systems selected  from spectroscopic surveys as those with the largest emission lines equivalent width (EW), e.g. EW(H$\beta) > $ 50\AA\  (or EW(H$\alpha) > $ 200\AA)  in their rest frame. 
The  lower limit in the EW of the recombination hydrogen lines   guarantees that a single and very young starburst,  less than $\sim 10 $Myr in age,  dominates the total luminosity output \citep[cf.][]{Dottori1981, Dottori1981b,  Leitherer1999, Melnick2000, Chavez2014}. As a result of this selection,  the possible contamination by an underlying older population or older clusters inside the spectrograph aperture as well as the ionizing photon escape are  minimised. In brief,  HIIG can be considered as ``naked" extremely young  starbursts. HIIG  thus selected are  spectroscopically  indistinguishable from young GHIIR in nearby galaxies (e.g.~30 Doradus or NGC~604 in M33). This is underlined by the fact that when first studied in detail, the prototypical\\
HIIGs I~Zw18 and II~Zw40 were dubbed  ``Isolated Extragalactic \hii\ regions" \citep{Sargent1970}.  

It is interesting to note that an important fraction of GHIIR and HIIG are complex systems with one massive cluster -- the youngest -- dominating the luminosity\footnote{These ionizing clusters are much more massive than the typical HII regions in our galaxy, like e.g. Orion.}. The possibility of these systems later merging and forming a single complex cluster is intriguing. 

From the analysis of the observed distribution of EW of the Balmer lines, \cite{Terlevich2004} concluded that the evolution of HIIG is consistent with a succession of short starbursts separated by quiescent periods and that, while the emission lines trace the properties of the present dominant burst, the underlying stellar continuum traces the whole star-formation history of the galaxy. 
Thus, observables like the EW of the Balmer lines that combine an emission line flux, i.e. a parameter pertaining to the present dominant burst, with the continuum flux, i.e. a parameter that traces the whole history of star formation, should not be used alone to characterize the present burst.

GHIIR  have long been applied to the calibration of the extragalactic distance scale \citep{Sandage1974} based on  a correlation between  region diameter and parent galaxy
luminosity.  A different  approach for using GHIIR as distance indicators was proposed by \citet{Melnick1977, Melnick1978}, who found  that their diameters are well correlated with the turbulent  widths of the nebular emission lines.  Melnick then showed that a
tighter correlation than the original one for GHIIR (diameter, parent galaxy luminosity) exists if one uses the mean turbulent velocity of the three largest HIIRs instead of their diameters.  It has since been clear that GHIIR can be used as distance indicators provided an adequate high quality calibration sample  is obtained.

\citet{Terlevich1981} found  evidence from a  small  sample --- the one available at the time --- that  scaling relations   that apply to gravitationally bound stellar systems like elliptical galaxies, bulges of spirals and galactic globular clusters,  apply also -- after taking evolution into account -- to HIIG and GHIIR. These are the relations between H$\beta$ luminosity, absolute blue magnitude, linear size, width of the emission lines and heavy element abundance. 
%It is now well established that GHIIR and HIIG exhibit a tight correlation between the luminosity and the width of their emission lines, the $L(\mathrm{H}\beta)-\sigma$ relation \citep{Terlevich1981}.  
The underlying physics of this relation is surprisingly simple: both the luminous and mechanical power of these objects, and the gravitational potential, stem from the same source: a young massive starburst. So the physical parameter at the basis of the correlation is the total mass of the burst component. Since HIIG have a very high luminosity per unit mass, and most of their luminosity is emitted in a few strong and narrow emission lines\footnote{HIIGs can reach $\ha $ luminosities of $10^{43}$ \ergs .}, the spectrum of these objects is easily observable out to large redshifts ($z > 3$) with present day instrumentation.

The scatter in the $L(\mathrm{H}\beta)-\sigma$  relation is small enough that it can be used to determine cosmic distances independently of redshift  \citep[see][]{Melnick1988,Melnick2000, Fuentes-Masip2000,Bosch2002, Siegel2005,Bordalo2011,Chavez2012,Chavez2014,FernandezArenas2018}.  
%HIIGs can reach $\hb $ luminosities larger than $10^{42}$ \ergs\ making them observable even at relatively  large redshifts ($z > 3$) with present-day  instrumentation. 

%{\bf \color{red} The fundamental plane  of elliptical galaxies evolves with redshift \citep{Treu2005,vanDokkum2007, Holden2010, Saglia2010, vandeSande2014, Zahid2015}; this  is attributed mainly to evolution in the mass-to-light (M/L) ratio of the stellar population but also  to some degree to dynamical evolution associated with the loss of mass from the system caused by stellar winds and SN ejecta.  Replacing the surface brightness with stellar mass surface density yields a stellar mass fundamental plane (MFP). Because stellar mass estimates ostensibly account for M/L ratio variations, changes in the MFP should trace the evolution of the structural or dynamical properties of galaxies. }

 %\ojo\ Remove? \ojo\ The FP topic is not mentioned at all in the rest of the paper \ojo\
 
The time is ripe to repeat  \citet{Terlevich1981} analysis, using now state-of-the-art data for GC, GHIIR, HIIG and numerical models for the dynamic and stellar evolution of the clusters and such is the purpose of this work. We will assume that these systems are massive virialized star clusters, and evolve them in time. 
We  refer to the massive ionizing stellar clusters as YMC and to the evolved non-ionizing  systems as super star clusters (SSC).
The paper is organized as follows: \S \ref{dataSa} introduces the data samples used; \S \ref{modellingEvo} presents the method to analyse the cluster evolution divided in  photometric evolution, mass loss, two-body relaxation and adiabatic expansion with the consequent changes in the velocity dispersion. \S \ref{evolMBsigma} analyses the evolution of the clusters  $M_B-\sigma$  plane. Discussion and conclusions are given in \S \ref{conclu}.

%============================= Section 2 =======================

\section{Data samples}\label{dataSa}

We selected from the literature the data samples to use for this exercise, as detailed below.

The globular clusters sample was obtained from \citet{Harris2010}\footnote{http://physwww.mcmaster.ca/$\sim$harris/mwgc.dat} a new revision of the 1996 version of the McMaster Catalogue, containing 157 objects, with parameters that now include central velocity dispersions.

The elliptical galaxies data were obtained from \citet{Faber1989}, a homogeneous and high quality photometric and spectroscopic catalogue of the nearest and brightest elliptical galaxies. Distance, central velocity dispersions, velocities relative to the cosmic rest frame and residual velocities relative to a velocity-flow model are calculated in their paper for individual galaxies and groups. 

The parameters for the spiral bulges are taken from \citet{Whitmore1979} who obtained  velocity dispersions  at the central regions of 30 galaxies, of which 21  are spirals. Their spectra  are from MacGraw-Hill Observatory using an intensified Reticon scanner.

\citet{Chilingarian2011} 	 present an analysis of high-resolution multi-object spectroscopic data for a sample of 24 Ultra Compact Dwarf galaxies in the  Fornax cluster obtained with
the ESO/VLT and the Fibre Large Array Multielement Spectrograph (FLAMES).

The compilation of ``local" HIIG (our own data) is from \citet{Chavez2014} who studied a sample of 128 HIIG (C14 hereinafter) selected from the SDSS on the basis of the equivalent width of  $\hb$  (EW($\hb$)$ >$ 50\AA ) and redshift z ($0.02<$z$<0.2$). The emission-line profiles were obtained with two different spectrographs: the High Dispersion Spectrograph (HDS) on Subaru, and the Ultraviolet and Visual Echelle Spectrograph (UVES) on the VLT; the photometric fluxes are from observations in 2m-class telescopes in Mexico: San Pedro M\'artir (SPM) in Baja California and Observatorio Astron\'omico Guillermo Haro (OAGH) in Cananea, Sonora. In addition \citet{Bordalo2011} published emission line widths and luminosities for 118 star forming regions in HIIG (BT11 from now on), obtaining the line widths  with the Fiber-fed Extended Range Optical Spectrograph (FEROS) installed on the ESO 2.2m telescope at La Silla Observatory in Chile, and the Coud\'e Spectrograph at Pico dos Dias Observatory (LNA/Brazil), while the spectrophotometry was obtained with the Boller \&\ Chivens Spectrograph on the ESO-La Silla 1.52m telescope \citep{Kehrig1-2004}. We have also included \citet{Melnick1988} 49 HIIG  high S/N spectrophotometric data that we will refer to as M88.  
The GHIIR include published data from \citet{Melnick1987} plus  new data  obtained in the Mexican 2m class telescopes mentioned above \citep{FernandezArenas2018}. 

We also included 13 high redshift (1.5<z<2.4)  HIIG obtained with MOSFIRE\@ Keck (Gonz\'alez-Mor\'an et al. in preparation) and 17 intermediate-z  (z\~0.7) HIIG  from  \cite{Hoyos2005} and  \cite{Perez2016} (hereinafter H05 and P16 respectively).

\begin{figure}
        \includegraphics[scale=0.4]{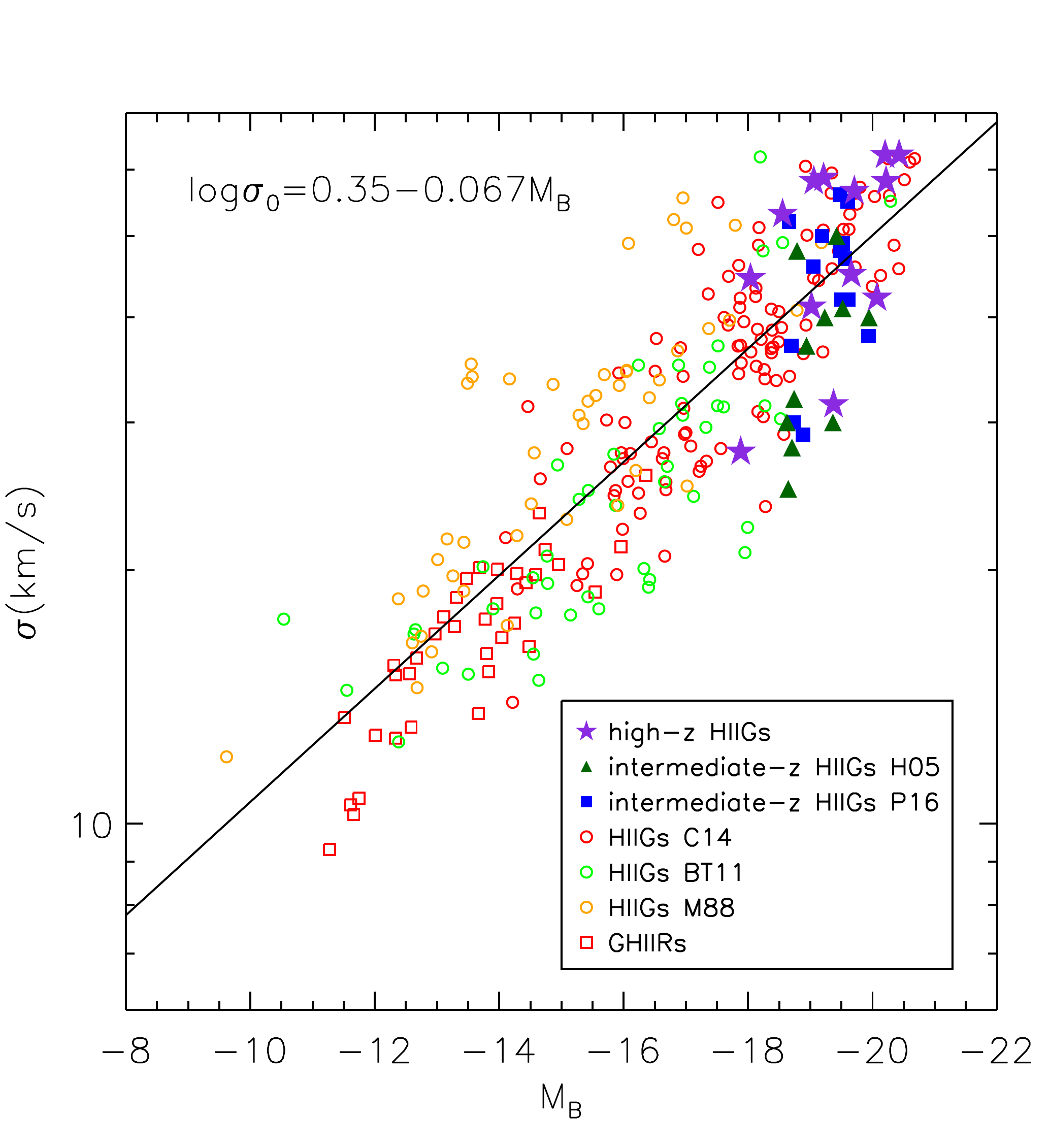}
        \caption{Correlation between the absolute magnitude in blue and the velocity dispersion for GHIIR and HIIG. The solid line represents the fit to the points, as indicated by the inset equation. The data sources are described in the text.}
        \label{fig:sigmaMb}
\end{figure}

The resulting $M_B-\sigma$ relation for the young systems, where $M_B$ was obtained  using equation (1)  is shown in figure~\ref{fig:sigmaMb}.

%\ojo\ A-L me dice que son 25... \ojo\ for 13 high-z and 17 intermediate-z \hii\ galaxies from Gonz\'alez-Mor\'an et al. (in preparation).

%As mentioned in the introduction, about 2/3 of the GHIIR and HIIG in our sample are complex systems with one cluster, usually the youngest, dominating the luminosity. The possibility of these systems later merging and forming a single complex cluster is consistent with the hierarchical merger scenario for the formation of massive clusters \cite[e.g.][]{Krumholz2007,Offner2008,Sabbi2012,Smilgys2017}

%============================= Section 3 =======================

\section{Modelling stellar cluster evolution}\label{modellingEvo}

In order to compare the  $(M_{B},\sigma)$ relation of HIIG and GHIIR  with that followed by the old speroidal stellar systems represented by  globular clusters, elliptical galaxies and bulges of spiral galaxies, we have evolved the relation $M_{B},\sigma$ for 12\,\gyr, taking into account the evolutionary changes in both magnitude and  velocity dispersion as we  discuss below.

 %%%%%%%%%%%%%%%% Figure 1 %%%%%%%%%%%%%%%%%%%

%Gieles and collaborators, in a series of papers \citep{Gieles2010,Gieles2011,Gieles2013} 
%discussed how the expansion of clusters is driven by mass loss (both from stellar evolution and from the dynamical effect of hard binaries and two body relaxation). In \citet{Gieles2010} they concluded that, because of the analytical expresion for relaxation time scales \citep{Spitzer1987} reproduced below, low mass systems are dynamically more evolved. As a result, if there was originally a Faber-Jackson type  relation between magnitude and size \citep{Faber1976}, 
 %for a cluster of total mass smaller than 10$^6$ \Msolar , such relation would have  been erased after about 10~\gyr\ of evolution.

All stellar systems expand as the result of stellar mass-loss.  For the smaller stellar systems in this paper, i.e. masses between $10^4$ and $10^6\, {\rm M}_\odot$ the effect of two-body relaxation results in a further expansion if the tidal field they are immersed in is relatively weak  \citep[e.g][]{Henon1965,Gieles2011}. 
 It is important to bare in mind that the ionising clusters of GHIIR and HIIG are much more massive than, e.g. the more common Orion  cluster and as we will show in the following sections, most of them will survive 12 \gyr\ of evolution without being disolved.
\citet{Gieles2010} showed that if globular clusters formed with a Faber-Jackson type relation \citep{Faber1976}, then the aforementioned effects would have moved the clusters with masses $\lesssim10^6\,{\rm M}_\odot$ away from this original relation. The fact that globular clusters are not following now the fundamental relations of galaxies does not exclude  therefore that they could have formed with the same relation.

\subsection{Photometric evolution}\label{PhotoEvo}
 
The absolute magnitude in B ($M_B$) was calculated from the observed equivalent widhts and $H\beta$ luminosities using  equation (6) from  \citet{Terlevich1981} modified with the new conversion from magnitudes to flux available at Gemini Observatory\footnote{Conversion from magnitudes to 
flux, http:
//www.gemini.edu/sciops/instruments/midir-resources/
imaging-calibrations/fluxmagnitude-conversion}.

 \begin{equation}
  M_B=-2.5\log\frac{L(H\beta)}{EW(H\beta)}+79.78
  \label{eq:MB}
 \end{equation}

 The  magnitude corrected by evolution ($M_{B_{*}}$) can be expressed as: 
 \begin{equation}
M_{B_{*}}=M_{B}+\Delta M_{B}
\label{eq:DeltaMB}
 \end{equation}
with the value of  $M_{B}$ deduced from the data and an EW($\hb$) of 50\AA{} set to put an upper limit to the age of 5 Myrs according to instantaneous burst mode  stellar population synthesis models\footnote {We have used \textit{starburst99} \citep{Leitherer1999}.}. Figure \ref{fig:ModelMasa} (a) shows the evolution of $M_B$ over $10^{10}$ yr of a \textit{starburst99}  model cluster of $10^6\msun$. Evolving the magnitude up to an age of 12\,\gyr, we obtain a change in  $M_{B}$  of $\Delta M_{B}=7.3$. Changing the stellar models introduces variations of less than 0.3 magnitudes.

  \subsection{Stellar mass loss}

 %\subsection{Mass loss}

\begin{figure*}
  \centering
  \subfloat[]{\label{test:1a}\includegraphics[height=90mm,width=90mm]{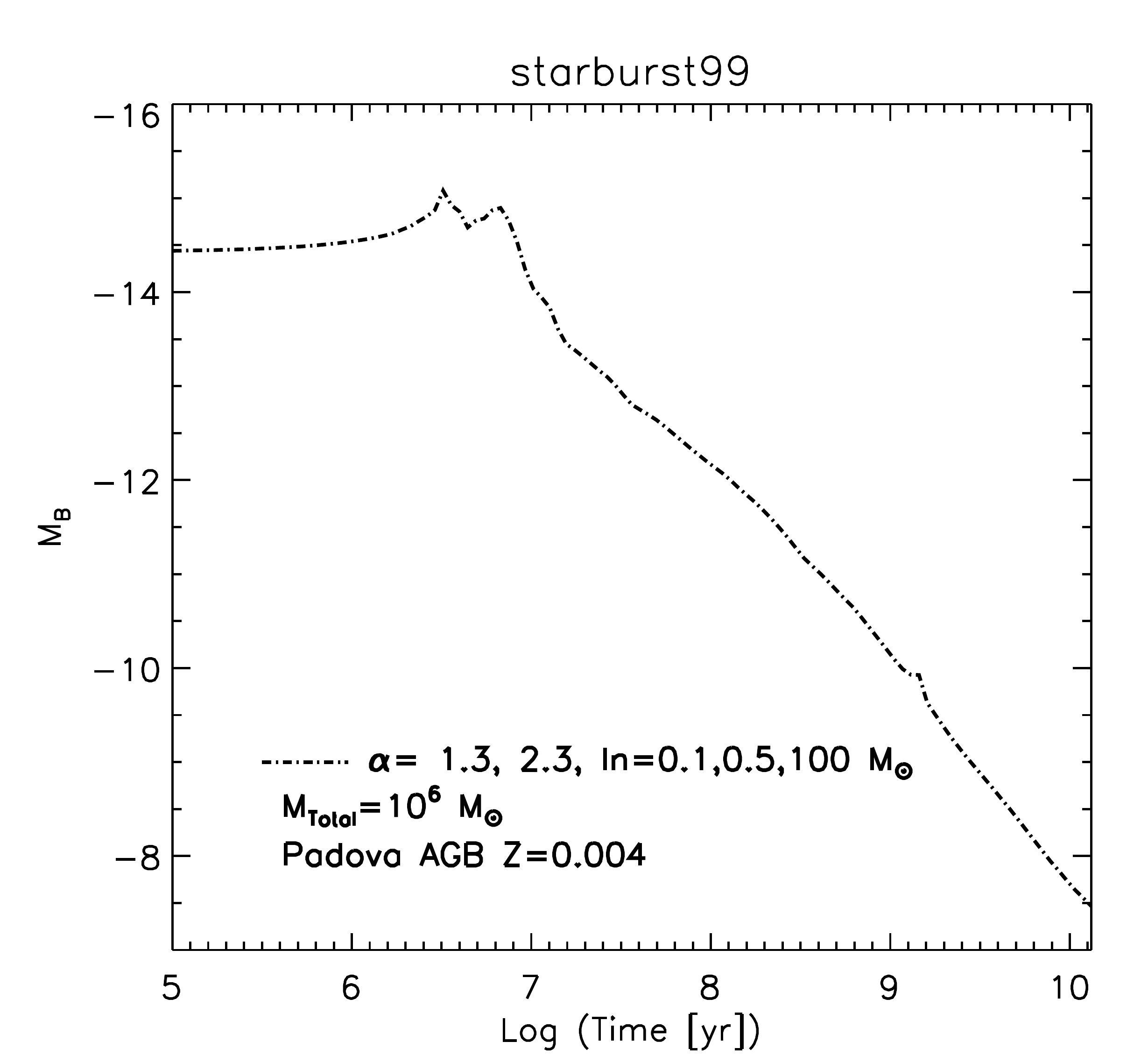}}
  \subfloat[]{\label{test:1b}\includegraphics[height=90mm,width=90mm]{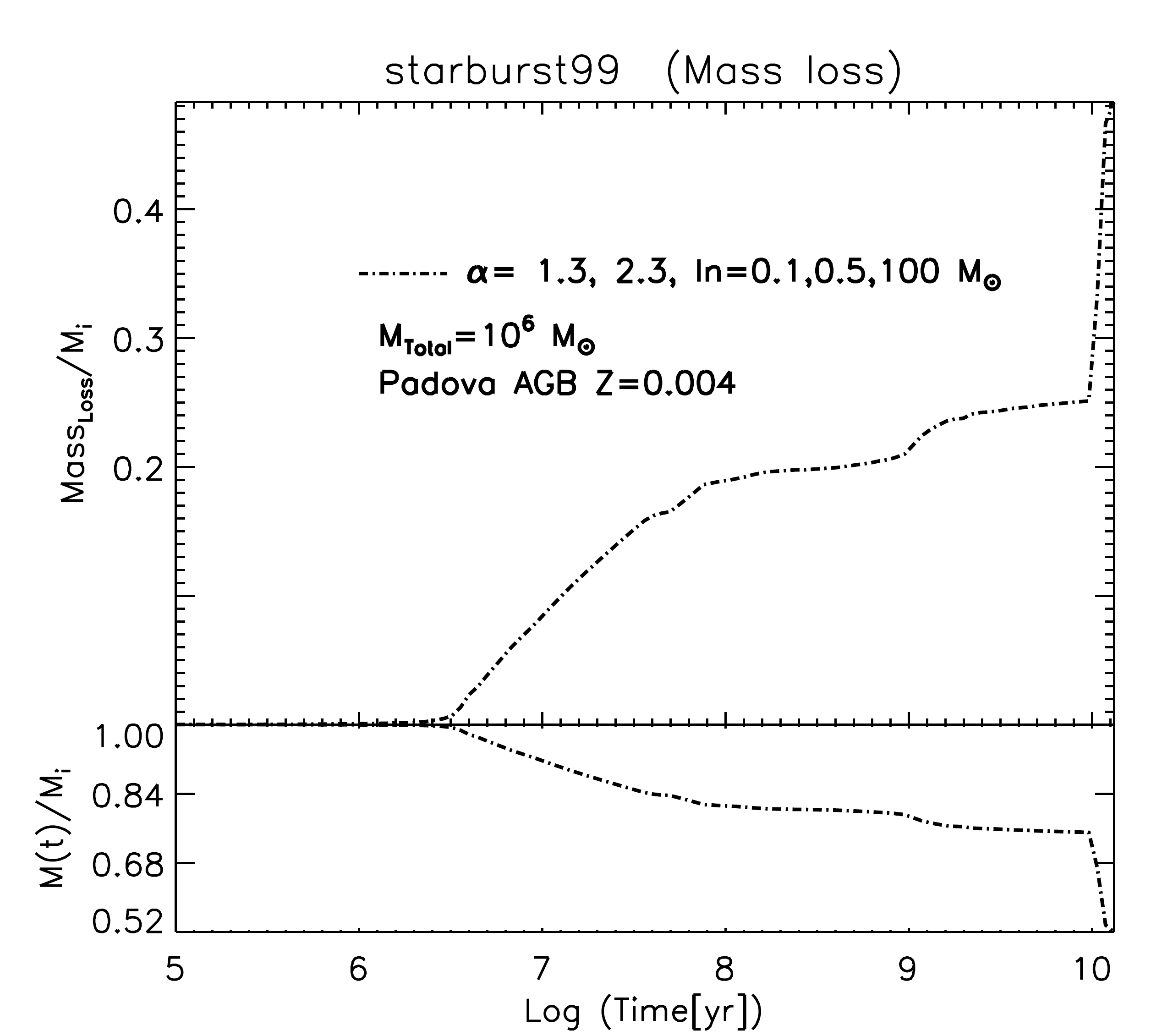}}
         \caption{(a): Evolution of  M$_B$ using \textit{starburst99} models as described in the text; (b): mass loss  due to stellar evolution from \textit{starburst99}, for a cluster with total mass 10$^{6}$\Msol,  Padova AGB tracks with Z=0.004 for a Kroupa IMF with slopes 1.3, 2.3. Fractional mass  loss (mass at time t relative to the initial mass) is represented at the bottom. 
         }
 \label{fig:ModelMasa}
\end{figure*}

\begin{figure}
  \centering
  {\label{test:2a}\includegraphics[height=90mm,width=90mm]{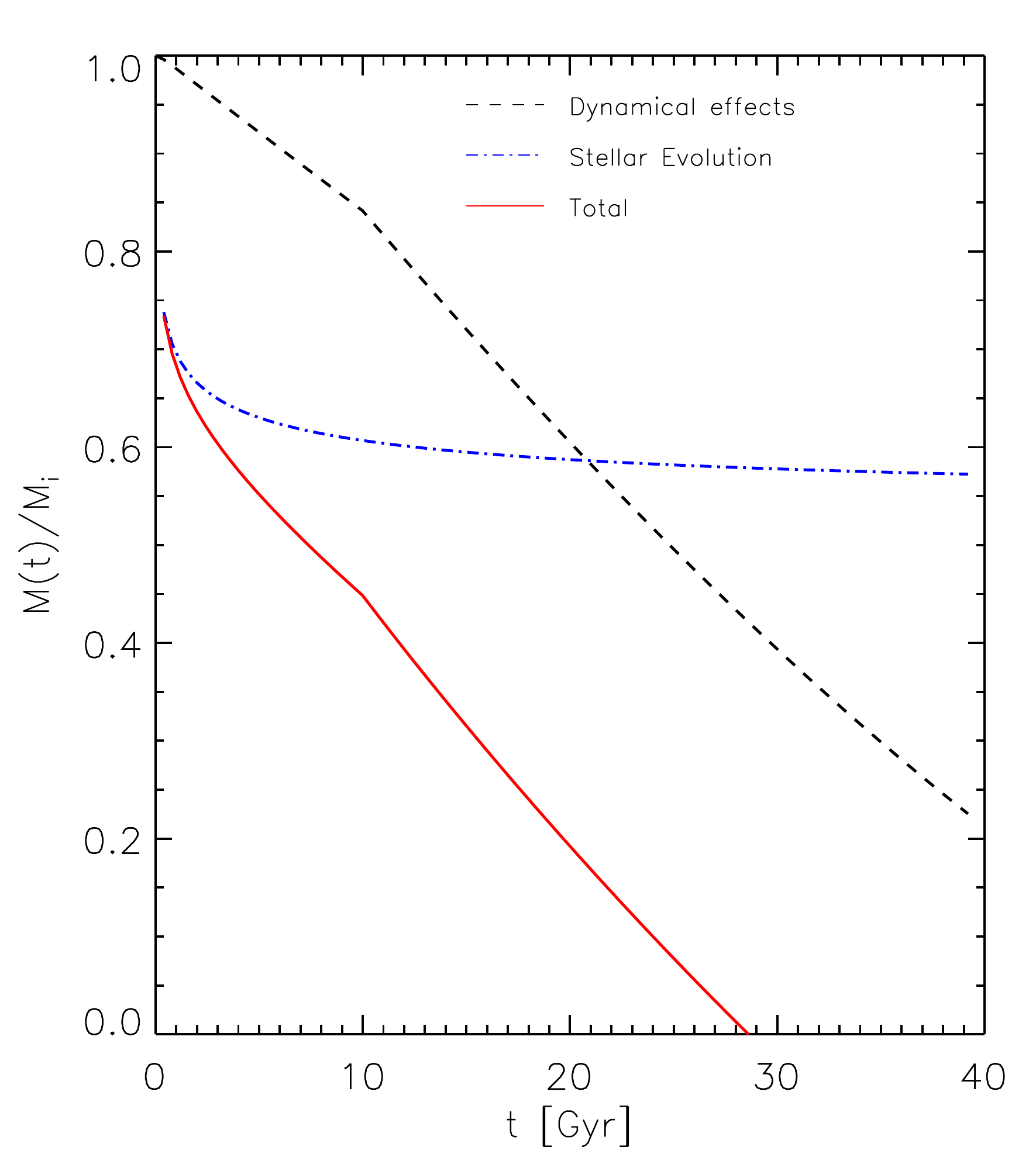}}
 \caption{ Change of mass due to dynamical and stellar evolution effects corresponding to model  ut10 from Lamers et al. (2010). }
  \label{fig:LamersGieles}
\end{figure}

When a gravitationally bound system loses mass, other parameters that characterize  the system like size and velocity dispersion are modified in consequence. Various processes generate loss of mass in the object: dynamical evolution and stellar evolution  (direct and induced) are the most important ones and they are protagonic  on different  time scales and at different stages of the evolution of the cluster \citep[see e.g. ][]{Lamers2010}. Figure \ref{fig:ModelMasa} (b) shows the mass lost due to stellar evolution by a  $10^6\msun$ cluster for a specific \textit{starburst99}  model whose parameters are given below and in the figure.
An example of the combined mass loss effect is shown in figure \ref{fig:LamersGieles}.  One could also add  the eventual loss of the remnant gas from the parental molecular cloud after the formation of the cluster. The caveat is that there is no gas in the simulations whose results  we are adopting as we will see in what follows. 

%To estimate the masses, we used a ratio $M/L=4.241$ in solar units of mass and luminosity, derived from \textit{starburst99}  for a 12 \gyr\ old stellar cluster and an instantaneous burst.  
We have obtained the ratio to the initial mass of  the mass lost by stellar evolution processes, using \textit{starburst99} models for a cluster with total mass 10$^{6}\msun$, Padova AGB tracks with metallicity Z=0.004 and  a Kroupa initial mass function with exponents 1.3, 2.3 and for lower, break and upper mass of 0.1, 0.5 and 100 $\msun$ respectively. We used  \citet{Lamers2010} models  to quantify the mass loss by dynamical effects, specifically their  uf10 model, one of their largest model clusters with     83,853~\Msolar\ and 128,000 stars, the smallest tidal field and Roche lobe underfilling and we are assuming the loss of molecular gas to be around 10\%. In table \ref{lostmass} we summarise the fraction of mass lost through each one of the  effects just mentioned.  Therefore the  change in mass gives $M_{f}/M_{i}=0.42$, and consequently the velocity dispersion using the virial theorem becomes  $\sigma_{f}=\sigma_{i}\sqrt{M_f/M_i}$.

%%%%%%%%%%%%%%%%%%%% Table 1 %%%%%%%%%%%%%%%% 
 \begin{table}
\centering
    \caption{\bf Lost mass fraction}\label{lostmass}
    \newcommand{\mc}[2]{\multicolumn{#1}{#2}}
 \begin{tabular}{lc}\\\hline
 \mc{1}{c}{$\Delta$Mass} & \mc{1}{c}{$M_{lost}/M_{i}$}\\\hline
 $\Delta M_{evolution}$ & 0.32\\
 $\Delta M_{dynamical}$ & 0.16\\
 $\Delta M_{mol-gas}$ & 0.10 \\
 $\Delta M_{lost-total}$ & 0.58\\\hline
 \end{tabular}
 \end{table}

 %\begin{center}
 %\label{massloss}
 %\begin{tabular}{ll}
% Mass & $M_{lost}/M_{i}$\\
% $M_{evolution}$ & 0.32\\
% $M_{dynamical}$ & 0.16\\
% $M_{gas}$ & 0.10 \\
% $M_{loss-total}$ & 0.58
 %\end{tabular}
% \end{center}

 %%%%%%%%%%%%%%%%%%%%%%%%%% Figure 2 %%%%%%%%%%%%%%
%%%%%%%%%%%%%%%%%%%%%%%%%%%%%%%%%%%%%%%%%%%%

%%%%%%%%%%%%%%%%%%%%%%%%%% Figure 3 %%%%%%%%%%%%%%
%%%%%%%%%%%%%%%%%%%%%%%%%%%%%%%%%%%%%%%%%%%%

%  \item Para calcular la masa se ha utilizado la relaciÃ³n $M/L= 4.241 $ en masas y luminodades solares. Derivada para un cÃºmulo con 12 giga-aÃ±os.

The mass loss in these systems drives an expansion, which in turn will also induce  a change in the velocity dispersion. To estimate such change  we use a model from \citet{Gieles2010}  for a 12\,\gyr\ old cluster. The change in velocity dispersion due to the expansion of the cluster is given by  $\sigma_{f}=\sigma_{i}\sqrt{\rh/\rhn}$, where ${\rhn/\rh}$ is the ratio of the initial to the final half-mass radius and  $\sigma_{f}$ and $\sigma_{i}$ are the final and initial velocity dispersions respectively.   

%The change of R$_h$ with mass for a cluster left to evolve for 12\,\gyr\ \citep{Gieles2010} is shown in the right panel of  figure \ref{fig:LamersGieles}.

% \begin{equation}
 %\frac{R_{h}}{R_{h0}}=\left[\left(\frac{T}{T_{*}}\right)^{2\delta}+{\left(\frac{\chi_{T}T}{T_{Rh0}}\right)^{4/3}}\right]^{1/2}
%\end{equation}
%\begin{equation}
%T_{Rh0}=0.138\frac{N^{1/2}R_{h0}^{3/2}}{G^{1/2}\bar{m}^{1/2}\ln\Lambda}
%\label{eq:DeltaRh0}
%\end{equation}
%
%\begin{equation}

 %\label{eq:DeltasigmaSize}
%\end{equation}

Finally combining the effects of mass loss and the  expansion of the cluster and always assuming that these systems are virialized, the velocity dispersion becomes:  

\begin{equation}
\sigma_{f}=\sigma_i\sqrt{\frac{M_f}{M_i}\frac{R_{h0}}{R_{h}}}
\label{eq:DeltasigmaTotal}
\end{equation}

%%%%%%%%%%%%%%%%%%%%% Section 4 %%%%%%%%%%%%
\vskip 5mm

\section{The Evolution of  YMC in the Mass$-\sigma$ plane}\label{evolMBsigma}

In this section we describe two  methods to evolve -- for a Hubble time -- the relation between  cluster mass $M$ and its 1D velocity dispersion $\sigma$ as the result of  mass loss due to the combined effect of dynamical and stellar evolution.

\subsection{Analytical relations}
We first consider the analytical description  of the evolution with time of the half-mass radius $\rh$ and $M$ taken from \citet{Gieles2010}. In this prescription the clusters are assumed to evolve in isolation (i.e. no tidal field).

\subsubsection{Expansion as the result of stellar mass-loss}
It is assumed that the mass decreases as the result of stellar mass loss as:

\begin{equation}
M(t)= M_0\left(\frac{t}{\ts}\right)^{-\delta}, \,\,\,t\ge\ts,\,\,\, \delta=0.07, \,\,\,\ts=2\,\myr.
\label{eq:mt}
\end{equation}
where $M_0$ is the initial mass at the time of cluster formation. This relation gives a ratio of $M/M_0\simeq0.54$  at an age of 12\,\gyr , which is in good agreement with most Single Stellar Population (SSP) models for a Kroupa like IMF \citep{Portegies2010}. 

In the early stages of evolution clusters expand adiabatically as the result of stellar mass-loss and in this regime the radius thus evolves as  \citep[e.g.][]{Hills1980} : 

\begin{equation}
\rh= \rhn\left(\frac{t}{\ts}\right)^{\delta}.
\label{eq:rt}
\end{equation}
The adiabatic expansion is a slow process  and gives a maximum increase of $\rh/\rhn\simeq2$ after a Hubble time.  

\subsubsection{Expansion as the result of two-body relaxation}

Two-body relaxation becomes important when the cluster age is comparable to the half-mass relaxation time-scale $\trh$:

\begin{equation}
\trh = 0.138\frac{N^{1/2}\rh^{3/2}}{G^{1/2}\bar{m}^{1/2}\ln\Lambda}.
\label{eq:trh}
\end{equation}
where $N$ is the number of stars in the system,   $G$ is the gravitational constant, $\bar{m}$ is the mean stellar mass and $\Lambda$ is the argument of the Coulomb logarithm that can take a value between 0.02N and 0.11N depending on the cluster mass function \citep{Giersz1996}.
% $\rh$ is the  half mass radius,

Isolated clusters then enter a self-similar expansion phase \citep{Henon1965} in which $\trh$ grows linearly with time and $\rh \propto t^{2/3}$ (in equation~\ref{eq:trh}). \citet{Gieles2010} propose a function that stitches together these two extremes in an attempt to match $\rh/\rhn$ for all values of $t/\trhn$

\begin{equation}
\rh(t)=\rhn\left(\left[\frac{t}{\ts}\right]^{2\delta}  + \left[\frac{\chifit t}{\trhn}\right]^{4/3}\right)^{1/2},  t\ge\ts.
\label{eq:rhtot}
\end{equation}
where $\chifit$ is  a time-dependent term and varies because of changes of the stellar mass function, and its value at an age $t$ is found  to be  well approximated by a simple power-law function

\begin{equation}
\chifit \approx 3\left(\frac{t}{\ts}\right)^{-0.3}, \,\,\,\ts\le t\lesssim20\,\gyr.
\label{eq:chifit}
\end{equation}
If we now define $\ts\equiv\min([2\,\myr,t])$ we have a continuous function for $\rh(\trhn,t)$, or $\rh(M_0,\rhn,t)$ for all $t$ \citep{Gieles2010}.

\subsubsection{Resulting $\sigma $ change}

We can then use the virial theorem to compute the 1D velocity dispersion as

\begin{equation}
\sigma(t) =\sqrt{\frac{0.4GM(t)}{3\rh(t)}}
\label{eq:sigmat}
\end{equation}

The result of this equation is shown as a black solid line at an age of 12\,\gyr\ in 
figure~\ref{msigma_plot} where we show (a) the evolution of the relation between the velocity dispersion $\sigma$ and the mass of the cluster  and (b) the relation between the radius and the mass. The initial  values are represented by the solid blue line. 
 
  %   \caption[ Evolution of the initial $M-\sigma$ relation]{Evolution of the clusters  in the Mass-velocity dispersion (a) and Mass-Radius (b)  planes. The initial values are represented by  the blue solid lines and filled circles and were computed for an initial half-mass density of $\rhoh=10^4\,\msun\,\pc^{-3}$.  The dashed lines show the results of the EMACSS simulations for an age of 12\,\gyr\ (see text).  Green triangles and black squares indicate a strong and weak tidal field respectively. The black solid line shows the analytical results of equations (\ref{eq:sigmat} and \ref{eq:rhtot}). The arrows join the initial and final values for each mass point. 

\vskip 5mm
\subsection{EMACSS models}

The analytical results of the previous section only apply to clusters evolving in isolation. To include the additional effect of a galactic tidal field, we use
the fast cluster evolution code Evolve Me A Cluster of StarS  
\citep[][EMACSS]{Alexander2012,Alexander2014}.
In this  code 
the evolution of the cluster is treated as the flow of energy within, as a fraction of the total energy per half mass relaxation time ($T_{\rm Rh}$) which can be calculated much faster than a full $N$-body simulation. Basically it consists of solving a series of coupled differential equations to provide the evolution of the cluster mass and radius, hence also that of the velocity dispersion.

We evolved clusters with initial masses between $3\times10^4  \msun $  and $3\times10^7  \msun $ in a tidal field with a singular isothermal halo with circular velocity of 220 \kms\ at galactocentric radii of $\rg=30\,\kpc $ with different initial conditions for the mass-radius relation including an initial half-mass density of $\rhoh=10^3, \,10^4,10^5\msun\,\pc^{-3}$. We use a Faber-Jackson initial mass-radius relation as  presented by \cite{Gieles2010}, and the derivation of a  mass-radius relation via L-$\sigma$ from \cite{FernandezArenas2018} described in the Appendix %\ref{iniCon}  
and with $R_h=(4/3)R_{eff}$ to correct for projection as in \citet{Gieles2010}.

\begin{figure*}
\centering
  \subfloat[]{\label{test:3a}\includegraphics[scale=0.35]{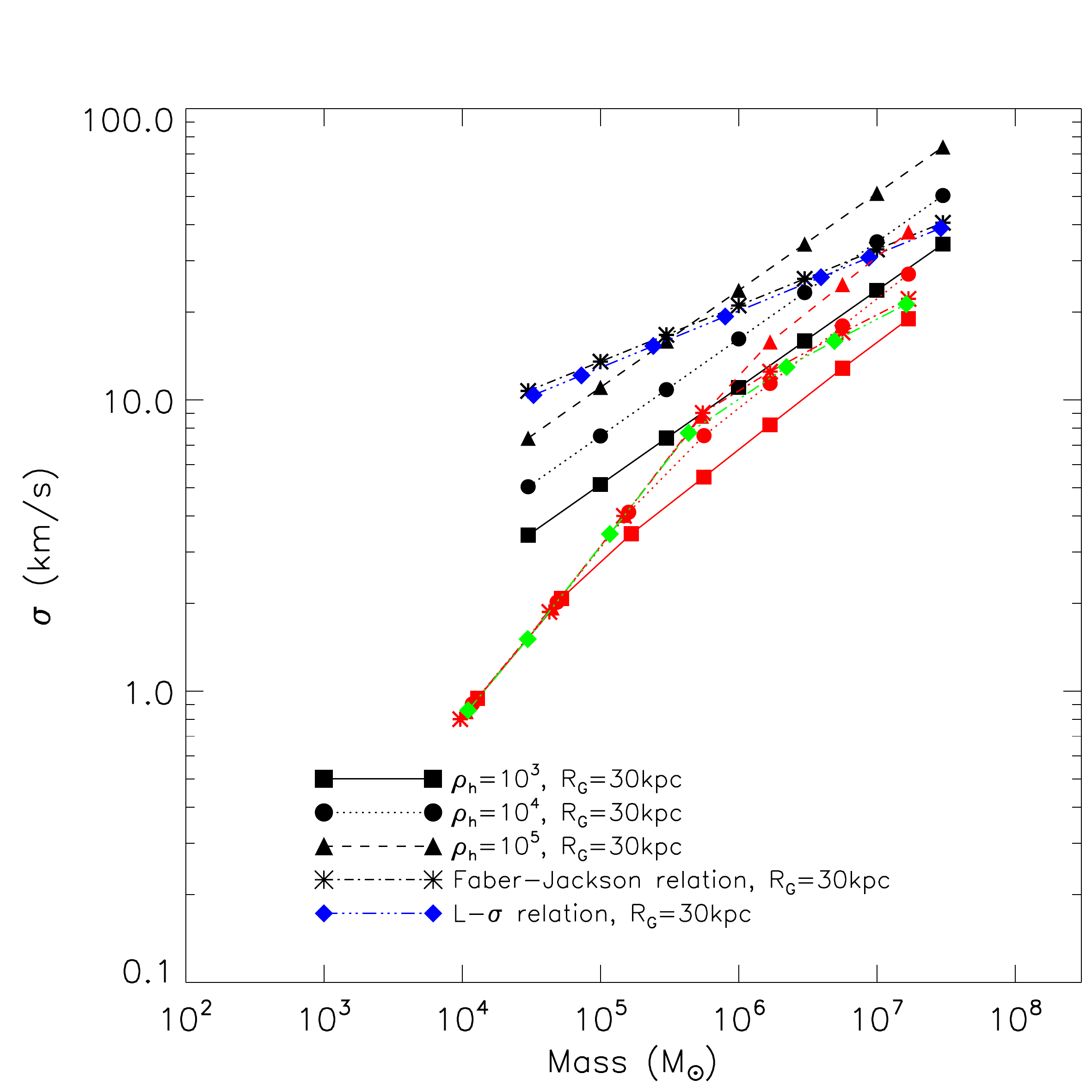}}
  \subfloat[]{\label{test:3b}\includegraphics[scale=0.35]{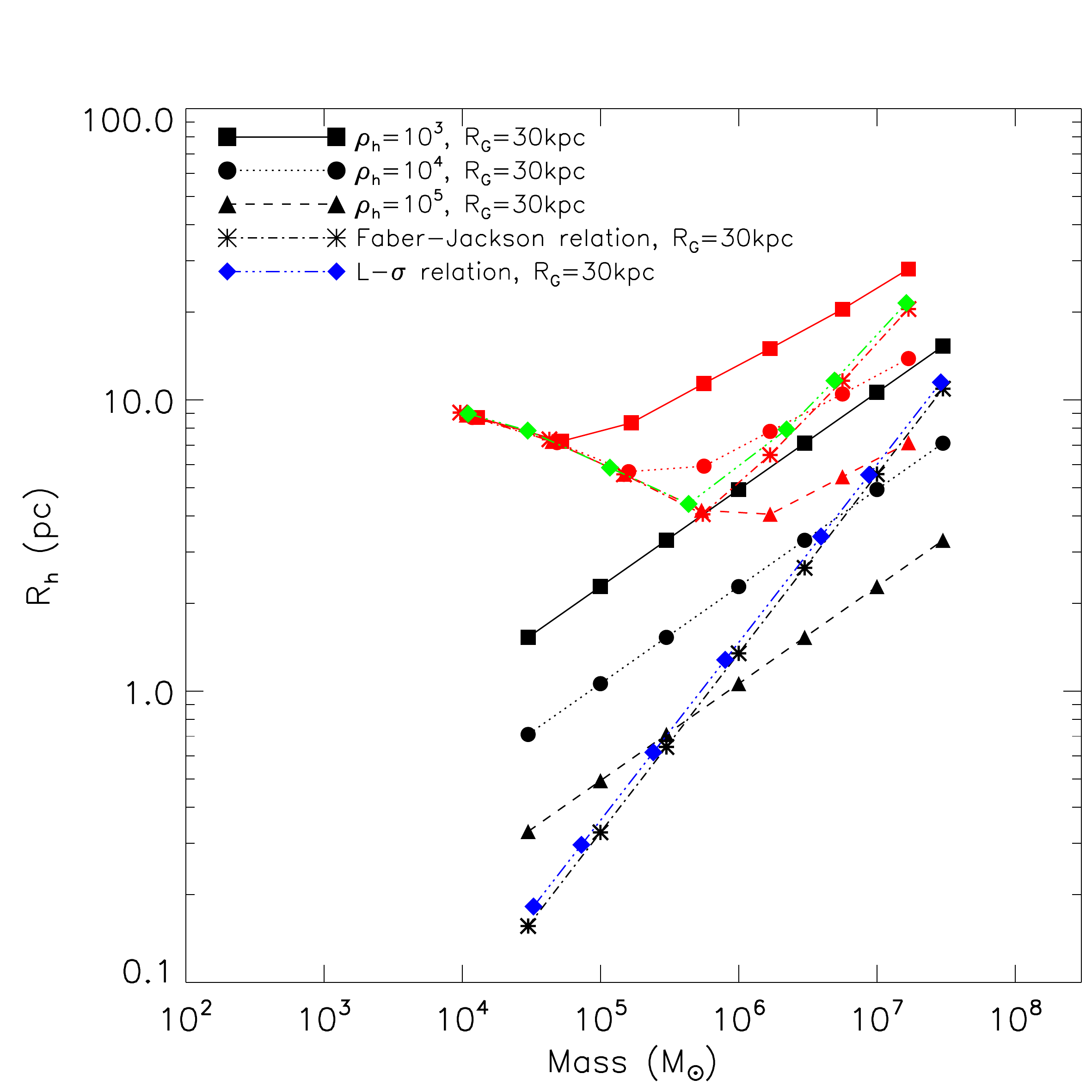}}
     \caption[ Evolution of the initial $M-\sigma$ relation]{Evolution of the clusters  (a):  in the Mass-$\sigma$  and (b): Mass-Radius  planes for three different initial conditions in a $\rg=30\,\kpc $. The black and blue points represent the intial conditions. The red and green points show the final results after 12\,\gyr\ of  evolution.      }
   \label{msigma_plot2}
\end{figure*}

The results of the models are shown in figure \ref{msigma_plot2}. It is noteworthy that the results of the initial conditions derived from the L-$\sigma$ relation agree surprisingly well with the estimate of \cite{Gieles2010} based on the  ''de-ageing" by 10\,\gyr\ of the observed Faber-Jackson relation and also with the Mass-Radius relation.

%%%%%%%%%%%%%% Figure 4 %%%%%%%%%%%%%%%

%We evolved clusters with initial masses between $3\times10^4  \msun $  and $3\times10^7  \msun $  in two different tidal fields: a singular isothermal halo with circular velocity of 220 \kms\  at galactocentric radii $\rg=3\,\kpc $ and $\rg=30\,\kpc $. The velocity dispersion was computed from $M$ and $\rh$ by EMACSS using equation~\ref{eq:sigmat}.

We also included simulations of the evolution of a  cluster in a strong tidal field changing the galactocentric radius to $\rg=3\,\kpc $. The velocity dispersion was computed from $M$ and $\rh$ by EMACSS using equation~\ref{eq:sigmat}.

\begin{figure*}
\centering
  \subfloat[]{\label{test:3a}\includegraphics[scale=0.35]{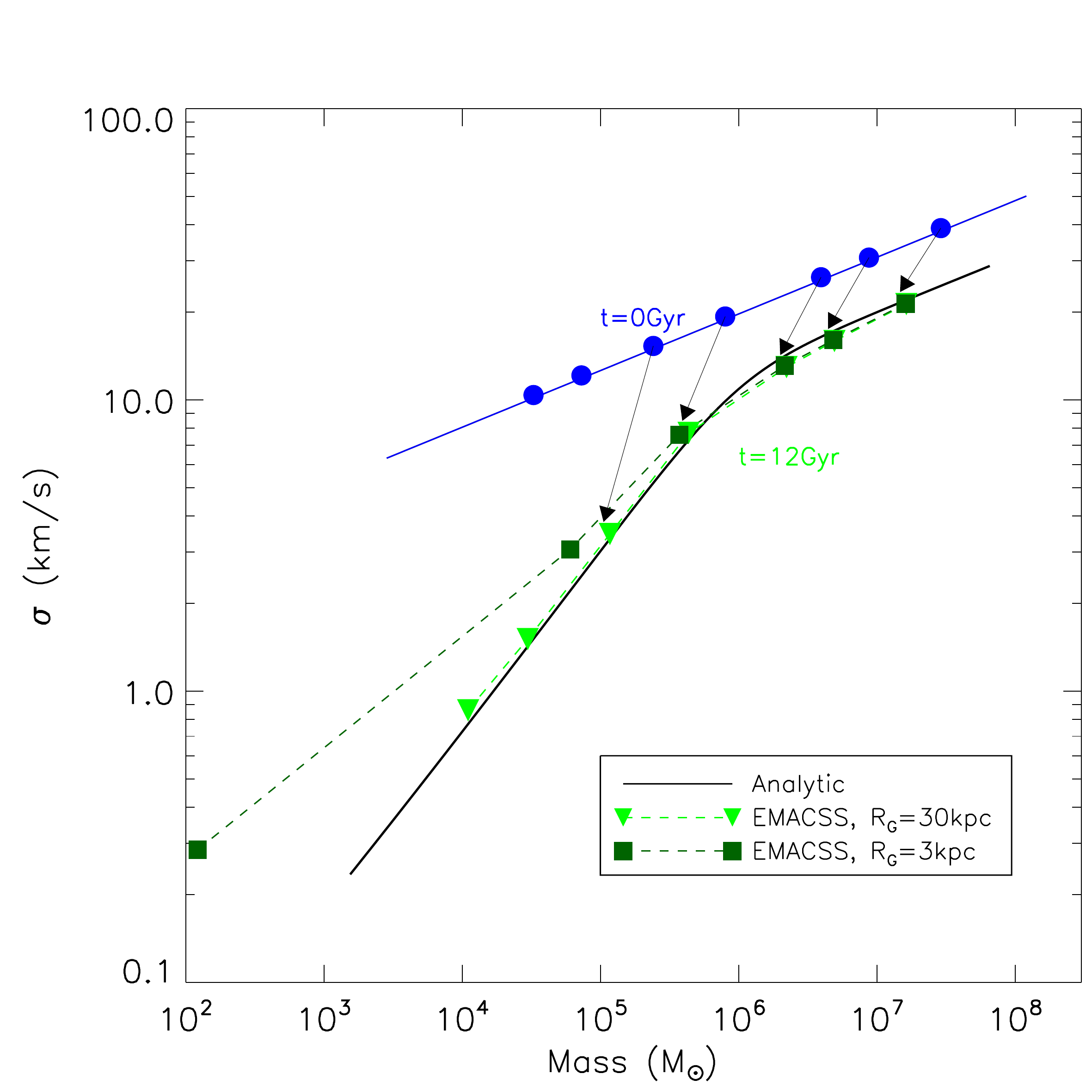}}
  \subfloat[]{\label{test:3b}\includegraphics[scale=0.35]{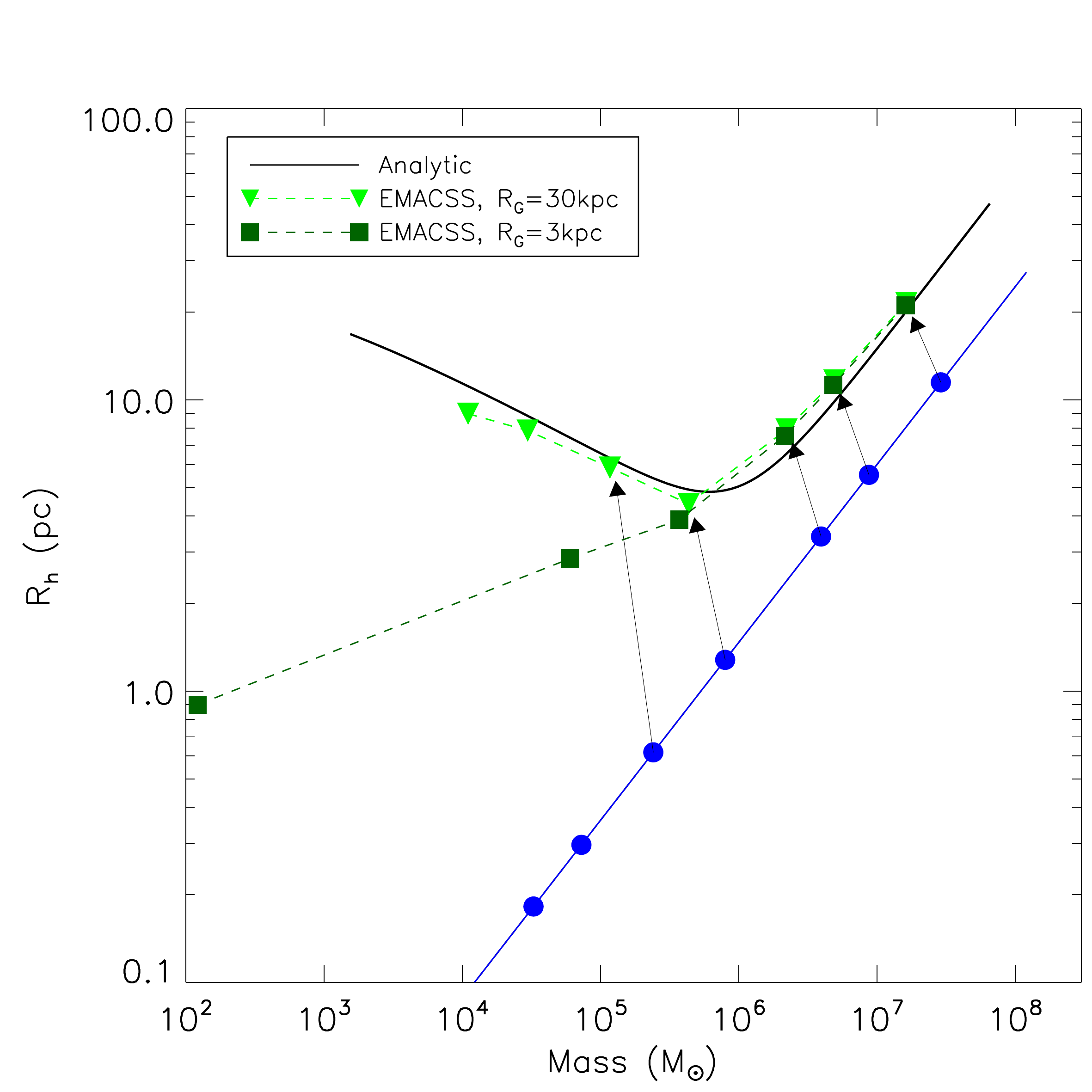}}
  
 %  \subfloat[]{\label{test:4a}\includegraphics[scale=0.3]{sigma_rho_fj.pdf}}
 %  \subfloat[]{\label{test:3b}\includegraphics[scale=0.3]{radio_rho_fj.pdf}}\\
   
 %   \subfloat[]{\label{test:5a}\includegraphics[scale=0.3]{sigma_rho_1e3.pdf}}
%  \subfloat[]{\label{test:3b}\includegraphics[scale=0.3]{radio_rho_1e3.pdf}}\\
    
%     \subfloat[]{\label{test:6a}\includegraphics[scale=0.3]{sigma_rho_1e4.pdf}}
 % \subfloat[]{\label{test:3b}\includegraphics[scale=0.3]{radio_rho_1e4.pdf}}\\
    
  %   \subfloat[]{\label{test:7a}\includegraphics[scale=0.3]{sigma_rho_1e5.pdf}}  
 %    \subfloat[]{\label{test:3b}\includegraphics[scale=0.3]{radio_rho_1e5.pdf}}\\
     \caption[ Evolution of the initial $M-\sigma$ relation]{Evolution of the clusters  in the Mass-velocity dispersion (a) and Mass-Radius (b)  planes. The initial values are represented by  the blue solid lines and filled circles (see Appendix \ref{iniCon}).  The dashed lines show the results of the EMACSS simulations for an age of 12\,\gyr\ (see text).  Green triangles and black squares indicate a strong and weak tidal field respectively. The black solid line shows the analytical results of equations (\ref{eq:sigmat} and \ref{eq:rhtot}). The arrows join the initial and final values for each mass point. 
     }
   \label{msigma_plot}
\end{figure*}

The results are shown as filled symbols in figure ~\ref{msigma_plot}. For clusters in a weak tidal field the result is very similar to that of the analytical estimate. For the clusters in a strong tidal field the final $M-\sigma$ relation at low masses is again that of clusters with a constant density, hence it is almost parallel to the initial relation.
hasta
%%%%%%%%%%%%%% Figure 5 %%%%%%%%%%%%%%%

The main conclusion  from the EMACSS evolution models is the same reached in the previous section, i.e.  that the mass lost through stellar winds and supernovae introduces a slow adiabatic expansion  of the clusters that lose about 40 \% of their initial mass of stars and all the remnant gas. In addition,  for those systems with masses below about $10^6 \msun$  stellar dynamical evolution becomes important. That introduces a break in the  mass-velocity dispersion relation at about $10^6 \msun$  (see figure \ref{msigma_plot}-(a)). This break is also clearly visible in the mass-size relation as seen in the right hand side of figure \ref{msigma_plot}. 

%%%%%%%%%%%%%%% Section 5 %%%%%%%%%%%%%%%
\section{Discussion and conclusions}\label{conclu}

%We have shown  how the change in mass of the SSC that ionize the GHIIR and HIIG  caused by the combined mass loss due to stellar evolution and by the loss of stars through the normal process of the clusters dynamical evolution, cause changes in the size and in the velocity dispersion of the cluster. 

%Based on the findings of the previous sections 
 We  have calculated the evolution of an individual YMC responsible for the ionization of  GHIIR and HIIG as it ages over 12\,\gyr. The almost identical results that we obtained using either a numerical simulation (EMACSS) or the analytical approximation, legitimises the  use of the analytical expressions  to estimate the evolution of the velocity dispersion from equation \ref{eq:DeltasigmaTotal}. The luminosity evolution was estimated using the results from Section \ref{PhotoEvo}. 

\begin{figure*}
 \includegraphics[width=15.5cm]{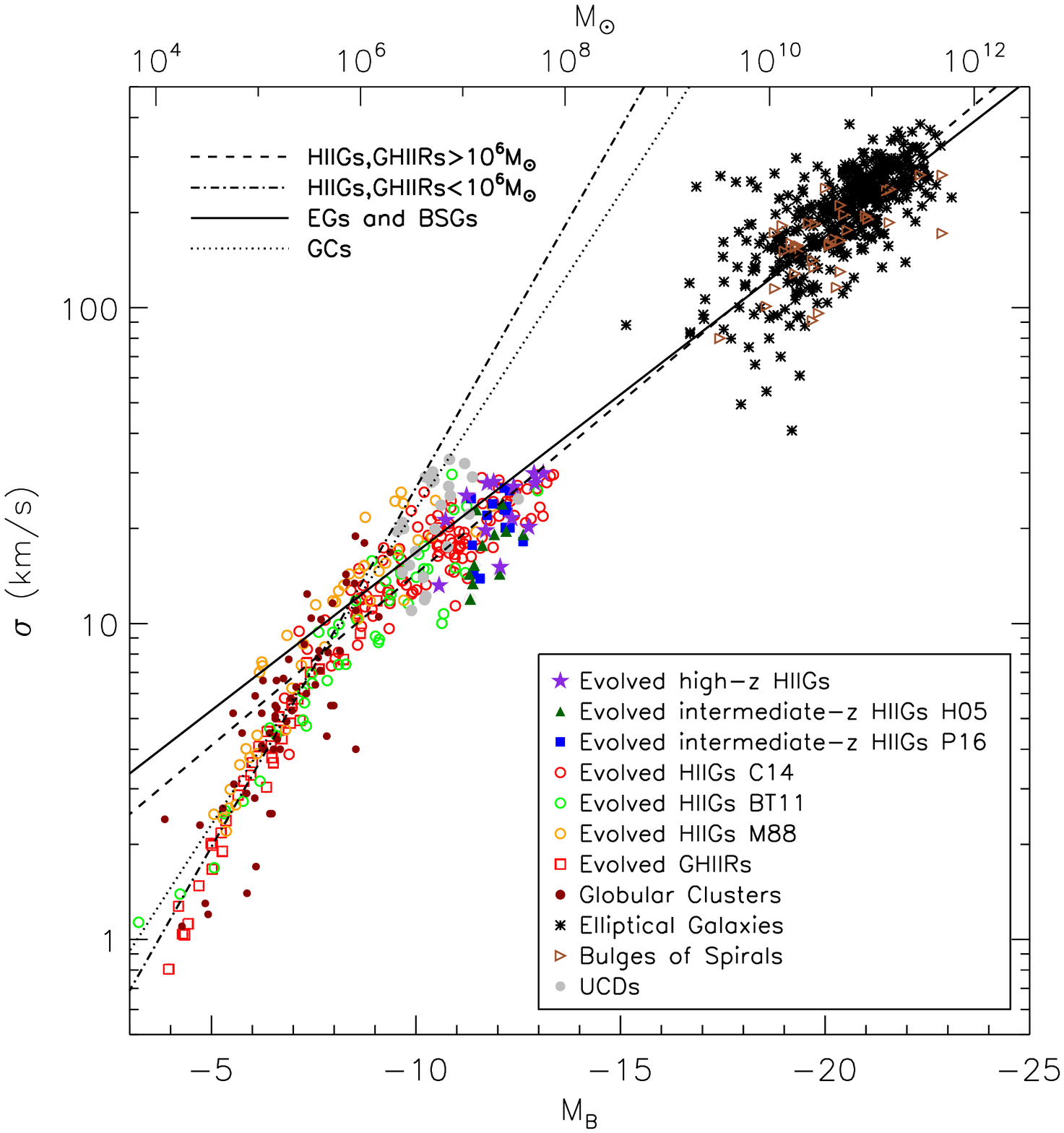}.
     \caption{ The position of evolved YMC is shown in the $\sigma$-M$_B$  plane on top of the  observed position of old spheroidal systems, i.e. GC, UCD, bulges of spiral galaxies (BSGs) and elliptical galaxies (EG). The solid line shows the fit to EG and BSGs while  the dotted line is the fit  to globular clusters given by \protect\cite{Hasegan2005} (equations 12 and 13). Also shown are the fits to the evolved YMC, the  
dashed line for those with $M>10^6 M\odot$, and the dot-dashed line, for those clusters with 
  $M<10^6 M\odot$ (equations (10), (11)). The upper axis labels represent the mass, estimated using an 
  M/L = 2 taken from \protect\cite{Baumgardt2017}.}
   \label{Ultimate_plot}
\end{figure*}

The position on the $\sigma \, vs. \, M_B$ plot of the YMC of our sample after 12\,\gyr\ of evolution is shown in figure 
 \ref{Ultimate_plot}. Also plotted in the figure and specified in the inset labels, are the data from our compilation as discussed in Section \ref{dataSa}.

We have performed fits to the evolved YMC separating the sample in two groups, those with $M<10^6 M\odot$ and those with $M>10^6 M\odot$. The resulting fits are:

\begin{equation}
\log\sigma=-(0.22M_B +0.85) \, \, M<10^6 M\odot
\end{equation}

\begin{equation}
\log\sigma=-(0.11M_B - 0.04) \, \, M>10^6 M\odot
\end{equation}\\
These fittings are shown in Figure \ref{Ultimate_plot}  as dot-dashed and dashed lines.

 \citet{Hasegan2005} investigated the scaling relations  for a compilation of galactic, M~31, and NGC~5128 GC, Local Group dSph, Virgo dE,N and their nuclei, Fornax Ultra Compact Dwarfs plus M~32. and elliptical galaxies. Their scaling relations converted to blue magnitudes are:

\begin{equation}
\log\sigma=-(0.20M_B + 0.64) \, \textrm{ for globular clusters}
\end{equation}

\begin{equation}
\log\sigma=-(0.10M_B  - 0.22)\, \textrm{ for elliptical galaxies}
\end{equation}\\

 The   fittings to  \citet{Hasegan2005} GC and ellipticals, the latter basically the Faber-Jackson relation, are also shown with solid lines in figure \ref{Ultimate_plot}.

%Figure \ref{Ultimate_plot} shows the position of the YMC of our sample after 12 Gyr of evolution. 

%Also reproduced in Figure \ref{Ultimate_plot} are the results from \citet{Hasegan2005}. That paper investigated the scaling relations  for a sample of galactic, M~31, and NGC~5128, globular clusters, Local Group dSph, Virgo dE,N and their nuclei, Fornax Ultra Compact Dwarfs plus M~32. Their compilation, including elliptical galaxies, is shown in Figure \ref{Ultimate_plot}, the continuos lines show their fittings to the globular cluster data (steeper line) and  ellipticals, basically the Faber-Jackson relation. 

It is quite striking that the initial position   of the YMC in the $\sigma$ vs. $M_B$ plane   in Figure \ref{fig:sigmaMb}, shift after 12\,\gyr\ of evolution to coincide with either the position of the globular clusters for YMC with initial mass $M < 10^6 \msun$ or with a continuation of the ellipticals, bulges of spirals and ultracompact dwarfs for the more massive YMC in figure \ref{Ultimate_plot}.

The two branches in the $M_B\, vs.\, \sigma$  plot are related to the break in the mass-sigma and mass-radius relations  seen in figure \ref{msigma_plot}-(a) and -(b) at about $10^6 \msun$ which manifest themselves as a 
change in the slope of the luminosity-sigma relation. This explains the observed different slopes in the L-$\sigma$ relation of elliptical galaxies and globular clusters. This change in slope is clearly seen in  the M$_B\, vs.\, \sigma$ plane in  figure \ref{Ultimate_plot}, in agreement with \cite{Hasegan2005} findings, also reproduced in the figure.

The remarkable result is that  not only the position of the evolved YMC in the  $\sigma$ vs. M$_B$ plane coincides with the positions of GC and Ultra Compact Dwarfs but also the position of the break in the relation is reproduced and the fits to the evolved YMC are strikingly similar to those of the \cite{Hasegan2005}  scaling relations for old spheroidal systems.

It is interesting to note that, our sample includes HIIG with redshifts up to  3.0 when the universe was about  2\,\gyr\ old and probably actively forming GCs. This is in line with  \cite{Elmegreen2012}  suggestion that the low metallicity halo globular clusters seen in spiral galaxies \citep{Brodie2006} could have been formed in dwarf galaxies, in particular Ly$\alpha$ emitters seen as young, metal poor  dwarf star-forming galaxies observed at high redshifts that are the building blocks of present day spirals.

%Regarding the abundance anomalies found in GC it is perhaps relevant that 

 As mentioned in the introduction, about $60\%$ of our sample of GHIIR and HIIG show evidence of either complex morphology or definite multiplicity with one cluster, the youngest, dominating the luminosity. The possibility of these systems later merging and forming a single complex cluster is consistent with the hierarchical merger scenario for the formation of massive clusters \cite[e.g.][]{Krumholz2007,Offner2008,Sabbi2012,Smilgys2017}.
\cite{Terlevich2004} concluded that the evolution of HIIG is consistent with a succession of starbursts separated by short quiescent periods and that, while the emission lines trace the properties of the present dominant burst, the underlying stellar continuum traces the whole star-formation history of the galaxy.
\citet{Sabbi2012} in their analysis  of the stellar population in the core of the prototypical GHIIR 30~Doradus (NGC~2070)  in the Large Magellanic Cloud identified two different stellar populations: the most massive and younger at the centre and a second one $\sim$1 Myr older at about 5pc away. The proximity and morphology of these clusters suggest that an ongoing merger may be occurring within the core of NGC~2070, a finding that is consistent with the predictions of models of hierarchical fragmentation of turbulent giant molecular clouds, according to which star clusters would be the final product of merging smaller sub-structures \citep{Bonnell2003, Bate2009, Federrath2010}.
Star formation will not be spread over the whole parent  molecular cloud, but will be localized in gravitationally bound pockets of gas \citep{Clark2005, Clark2008}.
Additional evidence for this scenario comes from the complexity of the structure of the other well studied GHIIR in the local group, NGC~604 in M~33 \citep{Maiz-Apellaniz2004}.
If GC form from mergers of smaller sub-systems, this may be related to the observed abundance
anomalies and perhaps also be 
the reason behind the high fraction of rotating GC. 

If the HII regions in our sample are indeed the progenitors of the old GCs observed in the nearby Universe, then they could potentially be used to shed light on the multiple population problem of (old) GCs. Globular clusters are characterised by light-element variations, in the form of broadened and/or multiple main sequences, believed to be the result of star-to-star helium variations, and Na-O and Mg-Al anticorrelations \citep[see][for a recent review]{BL18}. 
The origin of these abundance anomalies is topic of fierce debate, and various `polluters' have been put forward, such as asymptotic giant branch stars \citep[AGB stars, e.g.][]{2001ApJ...550L..65V}, (fast-rotating) massive stars \citep[$\gtrsim20\,{\rm M}_\odot$, e.g.][]{2007A&A...464.1029D, 2009A&A...507L...1D}, and supermassive stars \citep[SMS, M $\gtrsim10^3\,{\rm M}_\odot$, ][]{2014MNRAS.437L..21D, 2018MNRAS.478.2461G}. These multiple populations
 were once thought to be present only in the old GCs, but N spreads have recently been found in clusters as young as 2 Gyr in the Small and Large Magellanic Clouds \citep[e.g.][]{2018MNRAS.473.2688M}.
Measuring these abundance anomalies in the youngest massive clusters ($2\,$Myr) is challenging and upper limits on the Al abundance  in several YMCs have been established by \citet{2016MNRAS.460.1869C}.
Small spreads in Al have been found in young ($\sim10-40\,$Myr) clusters in the Antennae galaxies \citep{2017MNRAS.468.2482L}. It is therefore not clear, whether YMCs in the Local Universe form in the same way as their older counterparts. In the SMS scenario, the SMS may still be present after most of the gas has been evacuated \citep{2018MNRAS.478.2461G},
 hence the HII regions -- especially those at redshift $z \sim 3$ -- could be used to look for strong outflows from a SMS wind, enriched in Al, Na, and He. 

In conclusion, we interpret our results as a strong indication  that both young GHIIR and HIIG can evolve to form globular clusters and ultra compact dwarf ellipticals in about 12\,\gyr. These results make a strong case for the detailed study  of the YMC in nearby GHIIR and HIIG given that it can provide important clues to the formation and evolution of GCs.

%%%%%%%%%%%% Figure 5  %%%%%%%%%%%%%%%%%%

\section*{Acknowledgements}

The authors enjoyed clarifying discussions with Claus Leitherer 
%on the appropriate conditions for aplying \textit{starburst99}. 
and with Bruce Elmegreen. We thank an anonymous referee for suggestions that helped to improve the paper. RT, ET, DF-A, RC and ALG-M are grateful to
the Mexican research council (CONACYT) for supporting this research
under grants  CB-2005-01-49847, CB-2007-01-84746, 
CB-2008-103365-F and 263561, and studentships 262132 and 224117. MG acknowledges financial support
 from the Royal Society (University Research Fellowship) and the European Research Council (ERC stG-335936, CLUSTERS). DF-A  also acknowledges the hospitality of the Kavli Institute for Cosmology in Cambridge, where part of this work was completed.

%REFERENCES
\bibliographystyle{mnras}
\bibliography{bibpaper2018-08}

\begin{thebibliography}{}
\makeatletter
\relax
\def\mn@urlcharsother{\let\do\@makeother \do\$\do\&\do\#\do\^\do\_\do\%\do\~}
\def\mn@doi{\begingroup\mn@urlcharsother \@ifnextchar [ {\mn@doi@}
  {\mn@doi@[]}}
\def\mn@doi@[#1]#2{\def\@tempa{#1}\ifx\@tempa\@empty \href
  {http://dx.doi.org/#2} {doi:#2}\else \href {http://dx.doi.org/#2} {#1}\fi
  \endgroup}
\def\mn@eprint#1#2{\mn@eprint@#1:#2::\@nil}
\def\mn@eprint@arXiv#1{\href {http://arxiv.org/abs/#1} {{\tt arXiv:#1}}}
\def\mn@eprint@dblp#1{\href {http://dblp.uni-trier.de/rec/bibtex/#1.xml}
  {dblp:#1}}
\def\mn@eprint@#1:#2:#3:#4\@nil{\def\@tempa {#1}\def\@tempb {#2}\def\@tempc
  {#3}\ifx \@tempc \@empty \let \@tempc \@tempb \let \@tempb \@tempa \fi \ifx
  \@tempb \@empty \def\@tempb {arXiv}\fi \@ifundefined
  {mn@eprint@\@tempb}{\@tempb:\@tempc}{\expandafter \expandafter \csname
  mn@eprint@\@tempb\endcsname \expandafter{\@tempc}}}

\bibitem[\protect\citeauthoryear{{Alexander} \& {Gieles}}{{Alexander} \&
  {Gieles}}{2012}]{Alexander2012}
{Alexander} P.~E.~R.,  {Gieles} M.,  2012, \mn@doi [\mnras]
  {10.1111/j.1365-2966.2012.20867.x}, \href
  {http://adsabs.harvard.edu/abs/2012MNRAS.422.3415A} {422, 3415}

\bibitem[\protect\citeauthoryear{{Alexander}, {Gieles}, {Lamers}  \&
  {Baumgardt}}{{Alexander} et~al.}{2014}]{Alexander2014}
{Alexander} P.~E.~R.,  {Gieles} M.,  {Lamers} H.~J.~G.~L.~M.,   {Baumgardt} H.,
   2014, \mn@doi [\mnras] {10.1093/mnras/stu899}, \href
  {http://adsabs.harvard.edu/abs/2014MNRAS.442.1265A} {442, 1265}

\bibitem[\protect\citeauthoryear{{Bastian} \& {Lardo}}{{Bastian} \&
  {Lardo}}{2018}]{BL18}
{Bastian} N.,  {Lardo} C.,  2018, ARAA, arXiv:1712.01286, \href
  {http://adsabs.harvard.edu/abs/2017arXiv171201286B} {}

\bibitem[\protect\citeauthoryear{{Bate}}{{Bate}}{2009}]{Bate2009}
{Bate} M.~R.,  2009, \mn@doi [\mnras] {10.1111/j.1365-2966.2008.14106.x}, \href
  {http://adsabs.harvard.edu/abs/2009MNRAS.392..590B} {392, 590}

\bibitem[\protect\citeauthoryear{{Baumgardt}}{{Baumgardt}}{2017}]{Baumgardt2017}
{Baumgardt} H.,  2017, \mn@doi [\mnras] {10.1093/mnras/stw2488}, \href
  {http://adsabs.harvard.edu/abs/2017MNRAS.464.2174B} {464, 2174}

\bibitem[\protect\citeauthoryear{{Bonnell}, {Bate}  \& {Vine}}{{Bonnell}
  et~al.}{2003}]{Bonnell2003}
{Bonnell} I.~A.,  {Bate} M.~R.,   {Vine} S.~G.,  2003, \mn@doi [\mnras]
  {10.1046/j.1365-8711.2003.06687.x}, \href
  {http://adsabs.harvard.edu/abs/2003MNRAS.343..413B} {343, 413}

\bibitem[\protect\citeauthoryear{{Bordalo} \& {Telles}}{{Bordalo} \&
  {Telles}}{2011}]{Bordalo2011}
{Bordalo} V.,  {Telles} E.,  2011, \mn@doi [\apj] {10.1088/0004-637X/735/1/52},
  \href {http://adsabs.harvard.edu/abs/2011ApJ...735...52B} {735, 52}

\bibitem[\protect\citeauthoryear{{Bosch}, {Terlevich}  \& {Terlevich}}{{Bosch}
  et~al.}{2002}]{Bosch2002}
{Bosch} G.,  {Terlevich} E.,   {Terlevich} R.,  2002, \mn@doi [\mnras]
  {10.1046/j.1365-8711.2002.04967.x}, \href
  {http://adsabs.harvard.edu/abs/2002MNRAS.329..481B} {329, 481}

\bibitem[\protect\citeauthoryear{{Brodie} \& {Strader}}{{Brodie} \&
  {Strader}}{2006}]{Brodie2006}
{Brodie} J.~P.,  {Strader} J.,  2006, \mn@doi [\araa]
  {10.1146/annurev.astro.44.051905.092441}, \href
  {http://adsabs.harvard.edu/abs/2006ARA%26A..44..193B} {44, 193}

\bibitem[\protect\citeauthoryear{{Cabrera-Ziri}, {Lardo}, {Davies}, {Bastian},
  {Beccari}, {Larsen}  \& {Hernandez}}{{Cabrera-Ziri}
  et~al.}{2016}]{2016MNRAS.460.1869C}
{Cabrera-Ziri} I.,  {Lardo} C.,  {Davies} B.,  {Bastian} N.,  {Beccari} G.,
  {Larsen} S.~S.,   {Hernandez} S.,  2016, \mn@doi [\mnras]
  {10.1093/mnras/stw1090}, \href
  {http://adsabs.harvard.edu/abs/2016MNRAS.460.1869C} {460, 1869}

\bibitem[\protect\citeauthoryear{{Ch{\'a}vez}, {Terlevich}, {Terlevich},
  {Plionis}, {Bresolin}, {Basilakos}  \& {Melnick}}{{Ch{\'a}vez}
  et~al.}{2012}]{Chavez2012}
{Ch{\'a}vez} R.,  {Terlevich} E.,  {Terlevich} R.,  {Plionis} M.,  {Bresolin}
  F.,  {Basilakos} S.,   {Melnick} J.,  2012, \mn@doi [\mnras]
  {10.1111/j.1745-3933.2012.01299.x}, \href
  {http://adsabs.harvard.edu/abs/2012MNRAS.425L..56C} {425, L56}

\bibitem[\protect\citeauthoryear{{Ch{\'a}vez}, {Terlevich}, {Terlevich},
  {Bresolin}, {Melnick}, {Plionis}  \& {Basilakos}}{{Ch{\'a}vez}
  et~al.}{2014}]{Chavez2014}
{Ch{\'a}vez} R.,  {Terlevich} R.,  {Terlevich} E.,  {Bresolin} F.,  {Melnick}
  J.,  {Plionis} M.,   {Basilakos} S.,  2014, \mn@doi [\mnras]
  {10.1093/mnras/stu987}, \href
  {http://adsabs.harvard.edu/abs/2014MNRAS.442.3565C} {442, 3565}

\bibitem[\protect\citeauthoryear{{Chilingarian}, {Mieske}, {Hilker}  \&
  {Infante}}{{Chilingarian} et~al.}{2011}]{Chilingarian2011}
{Chilingarian} I.~V.,  {Mieske} S.,  {Hilker} M.,   {Infante} L.,  2011,
  \mn@doi [\mnras] {10.1111/j.1365-2966.2010.18000.x}, \href
  {http://adsabs.harvard.edu/abs/2011MNRAS.412.1627C} {412, 1627}

\bibitem[\protect\citeauthoryear{{Clark}, {Negueruela}, {Crowther}  \&
  {Goodwin}}{{Clark} et~al.}{2005}]{Clark2005}
{Clark} J.~S.,  {Negueruela} I.,  {Crowther} P.~A.,   {Goodwin} S.~P.,  2005,
  \mn@doi [\aap] {10.1051/0004-6361:20042413}, \href
  {http://adsabs.harvard.edu/abs/2005A%26A...434..949C} {434, 949}

\bibitem[\protect\citeauthoryear{{Clark}, {Muno}, {Negueruela}, {Dougherty},
  {Crowther}, {Goodwin}  \& {de Grijs}}{{Clark} et~al.}{2008}]{Clark2008}
{Clark} J.~S.,  {Muno} M.~P.,  {Negueruela} I.,  {Dougherty} S.~M.,  {Crowther}
  P.~A.,  {Goodwin} S.~P.,   {de Grijs} R.,  2008, \mn@doi [\aap]
  {10.1051/0004-6361:20077186}, \href
  {http://adsabs.harvard.edu/abs/2008A%26A...477..147C} {477, 147}

\bibitem[\protect\citeauthoryear{{Decressin}, {Meynet}, {Charbonnel},
  {Prantzos}  \& {Ekstr{\"o}m}}{{Decressin} et~al.}{2007}]{2007A&A...464.1029D}
{Decressin} T.,  {Meynet} G.,  {Charbonnel} C.,  {Prantzos} N.,   {Ekstr{\"o}m}
  S.,  2007, \mn@doi [\aap] {10.1051/0004-6361:20066013}, \href
  {http://adsabs.harvard.edu/abs/2007A%26A...464.1029D} {464, 1029}

\bibitem[\protect\citeauthoryear{{Denissenkov} \& {Hartwick}}{{Denissenkov} \&
  {Hartwick}}{2014}]{2014MNRAS.437L..21D}
{Denissenkov} P.~A.,  {Hartwick} F.~D.~A.,  2014, \mn@doi [\mnras]
  {10.1093/mnrasl/slt133}, \href
  {http://adsabs.harvard.edu/abs/2014MNRAS.437L..21D} {437, L21}

\bibitem[\protect\citeauthoryear{{Dottori}}{{Dottori}}{1981}]{Dottori1981}
{Dottori} H.~A.,  1981, \mn@doi [\apss] {10.1007/BF00652928}, \href
  {http://adsabs.harvard.edu/abs/1981Ap%26SS..80..267D} {80, 267}

\bibitem[\protect\citeauthoryear{{Dottori} \& {Bica}}{{Dottori} \&
  {Bica}}{1981}]{Dottori1981b}
{Dottori} H.~A.,  {Bica} E.~L.~D.,  1981, \aap, \href
  {http://adsabs.harvard.edu/abs/1981A%26A...102..245D} {102, 245}

\bibitem[\protect\citeauthoryear{{Elmegreen}, {Malhotra}  \&
  {Rhoads}}{{Elmegreen} et~al.}{2012}]{Elmegreen2012}
{Elmegreen} B.~G.,  {Malhotra} S.,   {Rhoads} J.,  2012, \mn@doi [\apj]
  {10.1088/0004-637X/757/1/9}, \href
  {http://adsabs.harvard.edu/abs/2012ApJ...757....9E} {757, 9}

\bibitem[\protect\citeauthoryear{{Esteban}, {Peimbert}, {Torres-Peimbert}  \&
  {Escalante}}{{Esteban} et~al.}{1998}]{Esteban98}
{Esteban} C.,  {Peimbert} M.,  {Torres-Peimbert} S.,   {Escalante} V.,  1998,
  \mn@doi [\mnras] {10.1046/j.1365-8711.1998.01335.x}, \href
  {http://adsabs.harvard.edu/abs/1998MNRAS.295..401E} {295, 401}

\bibitem[\protect\citeauthoryear{{Faber} \& {Jackson}}{{Faber} \&
  {Jackson}}{1976}]{Faber1976}
{Faber} S.~M.,  {Jackson} R.~E.,  1976, \mn@doi [\apj] {10.1086/154215}, \href
  {http://adsabs.harvard.edu/abs/1976ApJ...204..668F} {204, 668}

\bibitem[\protect\citeauthoryear{{Faber}, {Wegner}, {Burstein}, {Davies},
  {Dressler}, {Lynden-Bell}  \& {Terlevich}}{{Faber} et~al.}{1989}]{Faber1989}
{Faber} S.~M.,  {Wegner} G.,  {Burstein} D.,  {Davies} R.~L.,  {Dressler} A.,
  {Lynden-Bell} D.,   {Terlevich} R.~J.,  1989, \mn@doi [\apjs]
  {10.1086/191327}, \href {http://adsabs.harvard.edu/abs/1989ApJS...69..763F}
  {69, 763}

\bibitem[\protect\citeauthoryear{{Federrath}, {Banerjee}, {Clark}  \&
  {Klessen}}{{Federrath} et~al.}{2010}]{Federrath2010}
{Federrath} C.,  {Banerjee} R.,  {Clark} P.~C.,   {Klessen} R.~S.,  2010,
  \mn@doi [\apj] {10.1088/0004-637X/713/1/269}, \href
  {http://adsabs.harvard.edu/abs/2010ApJ...713..269F} {713, 269}

\bibitem[\protect\citeauthoryear{{Fern{\'a}ndez Arenas} et~al.,}{{Fern{\'a}ndez
  Arenas} et~al.}{2018}]{FernandezArenas2018}
{Fern{\'a}ndez Arenas} D.,  et~al., 2018, \mn@doi [\mnras]
  {10.1093/mnras/stx2710}, \href
  {http://adsabs.harvard.edu/abs/2018MNRAS.474.1250F} {474, 1250}

\bibitem[\protect\citeauthoryear{{Fuentes-Masip}, {Mu{\~n}oz-Tu{\~n}{\'o}n},
  {Casta{\~n}eda}  \& {Tenorio-Tagle}}{{Fuentes-Masip}
  et~al.}{2000}]{Fuentes-Masip2000}
{Fuentes-Masip} O.,  {Mu{\~n}oz-Tu{\~n}{\'o}n} C.,  {Casta{\~n}eda} H.~O.,
  {Tenorio-Tagle} G.,  2000, \mn@doi [\aj] {10.1086/301467}, \href
  {http://adsabs.harvard.edu/abs/2000AJ....120..752F} {120, 752}

\bibitem[\protect\citeauthoryear{{Gieles}, {Baumgardt}, {Heggie}  \&
  {Lamers}}{{Gieles} et~al.}{2010}]{Gieles2010}
{Gieles} M.,  {Baumgardt} H.,  {Heggie} D.~C.,   {Lamers} H.~J.~G.~L.~M.,
  2010, \mn@doi [\mnras] {10.1111/j.1745-3933.2010.00919.x}, \href
  {http://adsabs.harvard.edu/abs/2010MNRAS.408L..16G} {408, L16}

\bibitem[\protect\citeauthoryear{{Gieles}, {Heggie}  \& {Zhao}}{{Gieles}
  et~al.}{2011}]{Gieles2011}
{Gieles} M.,  {Heggie} D.~C.,   {Zhao} H.,  2011, \mn@doi [\mnras]
  {10.1111/j.1365-2966.2011.18320.x}, \href
  {http://adsabs.harvard.edu/abs/2011MNRAS.413.2509G} {413, 2509}

\bibitem[\protect\citeauthoryear{{Gieles} et~al.,}{{Gieles}
  et~al.}{2018}]{2018MNRAS.478.2461G}
{Gieles} M.,  et~al., 2018, \mn@doi [\mnras] {10.1093/mnras/sty1059}, \href
  {http://adsabs.harvard.edu/abs/2018MNRAS.478.2461G} {478, 2461}

\bibitem[\protect\citeauthoryear{{Giersz} \& {Heggie}}{{Giersz} \&
  {Heggie}}{1996}]{Giersz1996}
{Giersz} M.,  {Heggie} D.~C.,  1996, \mn@doi [\mnras]
  {10.1093/mnras/279.3.1037}, \href
  {http://adsabs.harvard.edu/abs/1996MNRAS.279.1037G} {279, 1037}

\bibitem[\protect\citeauthoryear{{Ha{\c s}egan} et~al.,}{{Ha{\c s}egan}
  et~al.}{2005}]{Hasegan2005}
{Ha{\c s}egan} M.,  et~al., 2005, \mn@doi [\apj] {10.1086/430342}, \href
  {http://adsabs.harvard.edu/abs/2005ApJ...627..203H} {627, 203}

\bibitem[\protect\citeauthoryear{{Harris}}{{Harris}}{2010}]{Harris2010}
{Harris} W.~E.,  2010, preprint, \href
  {http://adsabs.harvard.edu/abs/2010arXiv1012.3224H} {} (\mn@eprint {arXiv}
  {1012.3224})

\bibitem[\protect\citeauthoryear{{H{\'e}non}}{{H{\'e}non}}{1965}]{Henon1965}
{H{\'e}non} M.,  1965, Annales d'Astrophysique, \href
  {http://adsabs.harvard.edu/abs/1965AnAp...28...62H} {28, 62}

\bibitem[\protect\citeauthoryear{{Hills}}{{Hills}}{1980}]{Hills1980}
{Hills} J.~G.,  1980, \mn@doi [\apj] {10.1086/157703}, \href
  {http://adsabs.harvard.edu/cgi-bin/nph-bib_query?bibcode=1980ApJ...235..986H&db_key=AST}
  {235, 986}

\bibitem[\protect\citeauthoryear{{Holtzman} et~al.,}{{Holtzman}
  et~al.}{1992}]{Holtzman92}
{Holtzman} J.~A.,  et~al., 1992, \mn@doi [\aj] {10.1086/116094}, \href
  {http://adsabs.harvard.edu/abs/1992AJ....103..691H} {103, 691}

\bibitem[\protect\citeauthoryear{{Hoyos}, {Koo}, {Phillips}, {Willmer}  \&
  {Guhathakurta}}{{Hoyos} et~al.}{2005}]{Hoyos2005}
{Hoyos} C.,  {Koo} D.~C.,  {Phillips} A.~C.,  {Willmer} C.~N.~A.,
  {Guhathakurta} P.,  2005, \mn@doi [\apjl] {10.1086/499232}, \href
  {http://adsabs.harvard.edu/abs/2005ApJ...635L..21H} {635, L21}

\bibitem[\protect\citeauthoryear{{Kehrig}, {Telles}  \& {Cuisinier}}{{Kehrig}
  et~al.}{2004}]{Kehrig1-2004}
{Kehrig} C.,  {Telles} E.,   {Cuisinier} F.,  2004, \mn@doi [\aj]
  {10.1086/422922}, \href {http://adsabs.harvard.edu/abs/2004AJ....128.1141K}
  {128, 1141}

\bibitem[\protect\citeauthoryear{{Krumholz} \& {Bonnell}}{{Krumholz} \&
  {Bonnell}}{2007}]{Krumholz2007}
{Krumholz} M.~R.,  {Bonnell} I.~A.,  2007, preprint, \href
  {http://adsabs.harvard.edu/abs/2007arXiv0712.0828K} {} (\mn@eprint {arXiv}
  {0712.0828})

\bibitem[\protect\citeauthoryear{{Kunth} \& {{\"O}stlin}}{{Kunth} \&
  {{\"O}stlin}}{2000}]{Kunth2000}
{Kunth} D.,  {{\"O}stlin} G.,  2000, \mn@doi [\aapr] {10.1007/s001590000005},
  \href {http://adsabs.harvard.edu/abs/2000A%26ARv..10....1K} {10, 1}

\bibitem[\protect\citeauthoryear{{Lamers}, {Baumgardt}  \& {Gieles}}{{Lamers}
  et~al.}{2010}]{Lamers2010}
{Lamers} H.~J.~G.~L.~M.,  {Baumgardt} H.,   {Gieles} M.,  2010, \mn@doi
  [\mnras] {10.1111/j.1365-2966.2010.17309.x}, \href
  {http://adsabs.harvard.edu/abs/2010MNRAS.409..305L} {409, 305}

\bibitem[\protect\citeauthoryear{{Lardo}, {Cabrera-Ziri}, {Davies}  \&
  {Bastian}}{{Lardo} et~al.}{2017}]{2017MNRAS.468.2482L}
{Lardo} C.,  {Cabrera-Ziri} I.,  {Davies} B.,   {Bastian} N.,  2017, \mn@doi
  [\mnras] {10.1093/mnras/stx628}, \href
  {http://adsabs.harvard.edu/abs/2017MNRAS.468.2482L} {468, 2482}

\bibitem[\protect\citeauthoryear{{Leitherer} et~al.,}{{Leitherer}
  et~al.}{1999}]{Leitherer1999}
{Leitherer} C.,  et~al., 1999, \mn@doi [\apjs] {10.1086/313233}, \href
  {http://adsabs.harvard.edu/abs/1999ApJS..123....3L} {123, 3}

\bibitem[\protect\citeauthoryear{{Ma{\'{\i}}z-Apell{\'a}niz}, {P{\'e}rez}  \&
  {Mas-Hesse}}{{Ma{\'{\i}}z-Apell{\'a}niz} et~al.}{2004}]{Maiz-Apellaniz2004}
{Ma{\'{\i}}z-Apell{\'a}niz} J.,  {P{\'e}rez} E.,   {Mas-Hesse} J.~M.,  2004,
  \mn@doi [\aj] {10.1086/422925}, \href
  {http://adsabs.harvard.edu/abs/2004AJ....128.1196M} {128, 1196}

\bibitem[\protect\citeauthoryear{{Martocchia} et~al.,}{{Martocchia}
  et~al.}{2018}]{2018MNRAS.473.2688M}
{Martocchia} S.,  et~al., 2018, \mn@doi [\mnras] {10.1093/mnras/stx2556}, \href
  {http://adsabs.harvard.edu/abs/2018MNRAS.473.2688M} {473, 2688}

\bibitem[\protect\citeauthoryear{{Melnick}}{{Melnick}}{1977}]{Melnick1977}
{Melnick} J.,  1977, \mn@doi [\apj] {10.1086/155122}, \href
  {http://adsabs.harvard.edu/abs/1977ApJ...213...15M} {213, 15}

\bibitem[\protect\citeauthoryear{{Melnick}}{{Melnick}}{1978}]{Melnick1978}
{Melnick} J.,  1978, \aap, \href
  {http://adsabs.harvard.edu/abs/1978A%26A....70..157M} {70, 157}

\bibitem[\protect\citeauthoryear{{Melnick}, {Moles}, {Terlevich}  \&
  {Garcia-Pelayo}}{{Melnick} et~al.}{1987}]{Melnick1987}
{Melnick} J.,  {Moles} M.,  {Terlevich} R.,   {Garcia-Pelayo} J.-M.,  1987,
  \mnras, \href {http://adsabs.harvard.edu/abs/1987MNRAS.226..849M} {226, 849}

\bibitem[\protect\citeauthoryear{{Melnick}, {Terlevich}  \& {Moles}}{{Melnick}
  et~al.}{1988}]{Melnick1988}
{Melnick} J.,  {Terlevich} R.,   {Moles} M.,  1988, \mnras, \href
  {http://adsabs.harvard.edu/abs/1988MNRAS.235..297M} {235, 297}

\bibitem[\protect\citeauthoryear{{Melnick}, {Terlevich}  \&
  {Terlevich}}{{Melnick} et~al.}{2000}]{Melnick2000}
{Melnick} J.,  {Terlevich} R.,   {Terlevich} E.,  2000, \mn@doi [\mnras]
  {10.1046/j.1365-8711.2000.03112.x}, \href
  {http://adsabs.harvard.edu/abs/2000MNRAS.311..629M} {311, 629}

\bibitem[\protect\citeauthoryear{{Meylan}}{{Meylan}}{1993}]{Meylan1993}
{Meylan} G.,  1993, in {Smith} G.~H.,  {Brodie} J.~P.,  eds,  Astronomical
  Society of the Pacific Conference Series Vol. 48, The Globular Cluster-Galaxy
  Connection. p.~588

\bibitem[\protect\citeauthoryear{{Offner}, {Klein}  \& {McKee}}{{Offner}
  et~al.}{2008}]{Offner2008}
{Offner} S.~S.~R.,  {Klein} R.~I.,   {McKee} C.~F.,  2008, \mn@doi [\apj]
  {10.1086/590238}, \href {http://adsabs.harvard.edu/abs/2008ApJ...686.1174O}
  {686, 1174}

\bibitem[\protect\citeauthoryear{{P{\'e}rez}, {Hoyos}, {D{\'{\i}}az}, {Koo}  \&
  {Willmer}}{{P{\'e}rez} et~al.}{2016}]{Perez2016}
{P{\'e}rez} J.~M.,  {Hoyos} C.,  {D{\'{\i}}az} {\'A}.~I.,  {Koo} D.~C.,
  {Willmer} C.~N.~A.,  2016, \mn@doi [\mnras] {10.1093/mnras/stv1949}, \href
  {http://adsabs.harvard.edu/abs/2016MNRAS.455.3359P} {455, 3359}

\bibitem[\protect\citeauthoryear{{Portegies Zwart}, {McMillan}  \&
  {Gieles}}{{Portegies Zwart} et~al.}{2010}]{Portegies2010}
{Portegies Zwart} S.~F.,  {McMillan} S.~L.~W.,   {Gieles} M.,  2010, \mn@doi
  [\araa] {10.1146/annurev-astro-081309-130834}, \href
  {http://adsabs.harvard.edu/abs/2010ARA%26A..48..431P} {48, 431}

\bibitem[\protect\citeauthoryear{{Sabbi} et~al.,}{{Sabbi}
  et~al.}{2012}]{Sabbi2012}
{Sabbi} E.,  et~al., 2012, \mn@doi [\apjl] {10.1088/2041-8205/754/2/L37}, \href
  {http://adsabs.harvard.edu/abs/2012ApJ...754L..37S} {754, L37}

\bibitem[\protect\citeauthoryear{{Sandage} \& {Tammann}}{{Sandage} \&
  {Tammann}}{1974}]{Sandage1974}
{Sandage} A.,  {Tammann} G.~A.,  1974, \mn@doi [\apj] {10.1086/152906}, \href
  {http://adsabs.harvard.edu/abs/1974ApJ...190..525S} {190, 525}

\bibitem[\protect\citeauthoryear{{Sargent} \& {Searle}}{{Sargent} \&
  {Searle}}{1970}]{Sargent1970}
{Sargent} W.~L.~W.,  {Searle} L.,  1970, \mn@doi [\apjl] {10.1086/180644},
  \href {http://adsabs.harvard.edu/abs/1970ApJ...162L.155S} {162, L155}

\bibitem[\protect\citeauthoryear{{Shields} \& {Filippenko}}{{Shields} \&
  {Filippenko}}{1990}]{SF90}
{Shields} J.~C.,  {Filippenko} A.~V.,  1990, \mn@doi [\apjl] {10.1086/185695},
  \href {http://adsabs.harvard.edu/abs/1990ApJ...353L...7S} {353, L7}

\bibitem[\protect\citeauthoryear{{Siegel}, {Guzm{\'a}n}, {Gallego}, {Ordu{\~n}a
  L{\'o}pez}  \& {Rodr{\'{\i}}guez Hidalgo}}{{Siegel}
  et~al.}{2005}]{Siegel2005}
{Siegel} E.~R.,  {Guzm{\'a}n} R.,  {Gallego} J.~P.,  {Ordu{\~n}a L{\'o}pez} M.,
    {Rodr{\'{\i}}guez Hidalgo} P.,  2005, \mn@doi [\mnras]
  {10.1111/j.1365-2966.2004.08539.x}, \href
  {http://adsabs.harvard.edu/abs/2005MNRAS.356.1117S} {356, 1117}

\bibitem[\protect\citeauthoryear{{Smilgys} \& {Bonnell}}{{Smilgys} \&
  {Bonnell}}{2017}]{Smilgys2017}
{Smilgys} R.,  {Bonnell} I.~A.,  2017, \mn@doi [\mnras]
  {10.1093/mnras/stx2396}, \href
  {http://adsabs.harvard.edu/abs/2017MNRAS.472.4982S} {472, 4982}

\bibitem[\protect\citeauthoryear{{Terlevich} \& {Melnick}}{{Terlevich} \&
  {Melnick}}{1981}]{Terlevich1981}
{Terlevich} R.,  {Melnick} J.,  1981, \mnras, \href
  {http://adsabs.harvard.edu/abs/1981MNRAS.195..839T} {195, 839}

\bibitem[\protect\citeauthoryear{{Terlevich}, {Silich}, {Rosa-Gonz{\'a}lez}  \&
  {Terlevich}}{{Terlevich} et~al.}{2004}]{Terlevich2004}
{Terlevich} R.,  {Silich} S.,  {Rosa-Gonz{\'a}lez} D.,   {Terlevich} E.,  2004,
  \mn@doi [\mnras] {10.1111/j.1365-2966.2004.07432.x}, \href
  {http://adsabs.harvard.edu/abs/2004MNRAS.348.1191T} {348, 1191}

\bibitem[\protect\citeauthoryear{{Terlevich}, {Terlevich}, {Melnick},
  {Ch{\'a}vez}, {Plionis}, {Bresolin}  \& {Basilakos}}{{Terlevich}
  et~al.}{2015}]{Terlevich2015}
{Terlevich} R.,  {Terlevich} E.,  {Melnick} J.,  {Ch{\'a}vez} R.,  {Plionis}
  M.,  {Bresolin} F.,   {Basilakos} S.,  2015, \mn@doi [\mnras]
  {10.1093/mnras/stv1128}, \href
  {http://adsabs.harvard.edu/abs/2015MNRAS.451.3001T} {451, 3001}

\bibitem[\protect\citeauthoryear{{Terlevich} et~al.,}{{Terlevich}
  et~al.}{2016}]{Terlevich2016}
{Terlevich} R.,  et~al., 2016, \mn@doi [\aap] {10.1051/0004-6361/201628772},
  \href {http://adsabs.harvard.edu/abs/2016A%26A...592L...7T} {592, L7}

\bibitem[\protect\citeauthoryear{{Ventura}, {D'Antona}, {Mazzitelli}  \&
  {Gratton}}{{Ventura} et~al.}{2001}]{2001ApJ...550L..65V}
{Ventura} P.,  {D'Antona} F.,  {Mazzitelli} I.,   {Gratton} R.,  2001, \mn@doi
  [\apjl] {10.1086/319496}, \href
  {http://adsabs.harvard.edu/abs/2001ApJ...550L..65V} {550, L65}

\bibitem[\protect\citeauthoryear{{Whitmore}, {Schechter}  \&
  {Kirshner}}{{Whitmore} et~al.}{1979}]{Whitmore1979}
{Whitmore} B.~C.,  {Schechter} P.~L.,   {Kirshner} R.~P.,  1979, \mn@doi [\apj]
  {10.1086/157473}, \href {http://adsabs.harvard.edu/abs/1979ApJ...234...68W}
  {234, 68}

\bibitem[\protect\citeauthoryear{{de Mink}, {Pols}, {Langer}  \& {Izzard}}{{de
  Mink} et~al.}{2009}]{2009A&A...507L...1D}
{de Mink} S.~E.,  {Pols} O.~R.,  {Langer} N.,   {Izzard} R.~G.,  2009, \mn@doi
  [\aap] {10.1051/0004-6361/200913205}, \href
  {http://adsabs.harvard.edu/abs/2009A%26A...507L...1D} {507, L1}

\makeatother
\end{thebibliography}

\appendix
\section{Initial conditions estimation for the evolution calculations}\label{iniCon}

In the case of initial half-mass density the initial condition is computed by:
\begin{equation}
\rh=\left[\frac{3\mh}{8\pi\rhoh}\right]^{1/3}
\end{equation}

We also used the observe parameters of the ionising clusters or YMC in HIIG and GHIIR to estimate the total mass and derive their size from the virial theorem. 
From \citep{FernandezArenas2018} preferred solution (N13, Table 7) corrected for evolution to the median of EW(H$\beta)=100$ \AA , we get:

\begin{equation}\label{eqLS}
\log L(H\beta)=5.14\log\sigma+33.09
\end{equation}

From  \textit{starburst99} with Padova AGB tracks with metallicity Z=0.004 and  a Kroupa initial mass function, a cluster that has evolved reaching EW(H$\beta)=$100\AA{}   has:

\begin{equation}\label{eqLM}
\log L(H\beta)=\log(M_{tot})+33.73
\end{equation}
 
Using the eq. \ref{eqLM}  we can transform the eq. \ref{eqLS} to mass:

\begin{equation}\label{eqMT}
\log L(M_{tot})=5.14\log\sigma-0.64  \textrm{ or } M_{tot}=0.22\sigma^{5.14}
\end{equation}

From the virial theorem:

\begin{equation}\label{virial}
M_{tot}=233\eta R_{eff} \sigma^{2}
\end{equation}

Combining the eq. \ref{virial} and eq. \ref{eqMT} and assuming a mass profile following a King's law, we obtain:

\begin{equation}\label{reff}
R_{eff}=9.44\times10^{-5}\sigma^{3.14}
\end{equation}

The total cluster mass in equation \ref{eqMT} depends on assumptions about the IMF and stellar evolution tracks. The size estimate in equation \ref{reff} depends on the assumption that these YMC are virialized and follow a  King's mass profile. 
To compare with \citet{Gieles2010} mass-radius relation, we used a   $R_h=(4/3)R_{eff}$ that corrects for projection effects.
% The results agrees surprisingly well with the estimate of \cite{Gieles2010} based on the estimated ''de-ageing" by 10Gyrs of the observed Faber-Jacskon relation and Mass-Radius relation.

\bsp

\end{document}